\documentclass[draft]{agujournal2019}
\usepackage{url} %this package should fix any errors with URLs in refs.
\usepackage{lineno}
\usepackage{amsmath}
\usepackage[finalnew]{trackchanges} %for better track changes. finalnew option will compile document with changes incorporated.
\usepackage{soul}
\definecolor{c1}{RGB}{175, 207, 238}
\sethlcolor{white}
\draftfalse

\journalname{JGR: Planets}

\DeclareMathOperator{\atantwo}{atan2}

\begin{document}

\newcommand{\oc}{\mathrm{oc}}
\newcommand{\res}{\mathrm{res}}
\newcommand{\prim}{\mathrm{prim}}
\newcommand{\ind}{\mathrm{ind}}
\newcommand{\Jup}{\mathrm{J}}

\newcommand{\vecB}{\mathbf{B}}
\newcommand{\vecr}{\mathbf{r}}

%%%%%%%%%%%%%%%%%%%%%%%%%%%%%%%%%%%%%%%%%%%%%%%
%  TITLE
%
% (A title should be specific, informative, and brief. Use
% abbreviations only if they are defined in the abstract. Titles that
% start with general keywords then specific terms are optimized in
% searches)
%
%%%%%%%%%%%%%%%%%%%%%%%%%%%%%%%%%%%%%%%%%%%%%%%

% Example: \title{This is a test title}

\title{Detectability of Local Water Reservoirs in Europa's Surface Layer under Consideration of Coupled Induction}

%%%%%%%%%%%%%%%%%%%%%%%%%%%%%%%%%%%%%%%%%%%%%%%
%
%  AUTHORS AND AFFILIATIONS
%
%%%%%%%%%%%%%%%%%%%%%%%%%%%%%%%%%%%%%%%%%%%%%%%

% Authors are individuals who have significantly contributed to the
% research and preparation of the article. Group authors are allowed, if
% each author in the group is separately identified in an appendix.)

% List authors by first name or initial followed by last name and
% separated by commas. Use \affil{} to number affiliations, and
% \thanks{} for author notes.
% Additional author notes should be indicated with \thanks{} (for
% example, for current addresses).

\authors{J. Winkenstern\affil{1}, J. Saur\affil{1}}

\affiliation{1}{Institute of Geophysics and Meteorology, University of Cologne, Cologne, Germany}

\correspondingauthor{Jason Winkenstern}{jwinken3@uni-koeln.de}

%%%%%%%%%%%%%%%%%%%%%%%%%%%%%%%%%%%%%%%%%%%%%%%
% KEY POINTS
%%%%%%%%%%%%%%%%%%%%%%%%%%%%%%%%%%%%%%%%%%%%%%%
%  List up to three key points (at least one is required)
%  Key Points summarize the main points and conclusions of the article
%  Each must be 140 characters or fewer with no special characters or punctuation and must be complete sentences

% Example:
% \begin{keypoints}
% \item	List up to three key points (at least one is required)
% \item	Key Points summarize the main points and conclusions of the article
% \item	Each must be 140 characters or fewer with no special characters or punctuation and must be complete sentences
% \end{keypoints}

\begin{keypoints}
\item An analytical model for the mutual self-consistent induction of an ocean and a local, spherical water reservoir has been developed
\item During a flyby at 25 km altitude, liquid water reservoirs within Europa's icy crust cannot be detected with magnetometer measurements
\item A reservoir can be detected with one magnetometer on the surface directly above the reservoir and a second surface magnetometer nearby
\end{keypoints}

%%%%%%%%%%%%%%%%%%%%%%%%%%%%%%%%%%%%%%%%%%%%%%%
%
%  ABSTRACT and PLAIN LANGUAGE SUMMARY
%
% A good Abstract will begin with a short description of the problem
% being addressed, briefly describe the new data or analyses, then
% briefly states the main conclusion(s) and how they are supported and
% uncertainties.

% The Plain Language Summary should be written for a broad audience,
% including journalists and the science-interested public, that will not have 
% a background in your field.
%
% A Plain Language Summary is required in GRL, JGR: Planets, JGR: Biogeosciences,
% JGR: Oceans, G-Cubed, Reviews of Geophysics, and JAMES.
% see http://sharingscience.agu.org/creating-plain-language-summary/)
%
%%%%%%%%%%%%%%%%%%%%%%%%%%%%%%%%%%%%%%%%%%%%%%%

\begin{abstract}
The icy moon Europa is a primary target for the study of ocean worlds. Its subsurface ocean is expected to be subject to asymmetries on global scales (tidal deformation) and local scales (chaos regions, fractures). Here, we investigate the possibility to magnetic sound local asymmetries by calculating the induced magnetic fields generated by a radially symmetric ocean and a small, spherical water reservoir between the ocean and Europa's surface. The consideration of two conductive bodies introduces non-linear magnetic field coupling between them. We construct an analytical model to describe the coupling between two conductive bodies and calculate the induced fields within the parameter space of possible conductivity values and icy crust thicknesses. Given the plasma magnetic field perturbations, we find that a reservoir cannot be detected during a flyby at 25 km altitude using electromagnetic induction. Potential detection of liquid water reservoirs can be achieved by deploying magnetometers on Europa's surface, where one magnetometer is placed directly on the target region of interest and a second one in the nearby vicinity as reference to distinguish from global asymmetries. With this method, the smallest reservoir that can be detected has a radius of 8 km and a conductivity of 30 S/m. Larger reservoirs are resolvable at lower conductivities, with a 20 km reservoir requiring a conductivity of approximately 5 S/m.
\end{abstract}

\section*{Plain Language Summary}
Jupiter's icy moon Europa most likely harbors an ocean of liquid salt-water underneath its icy surface. In addition, small reservoirs of liquid water might exist within Europa's icy crust at very shallow depths. Such features offer interesting target regions for future lander missions in the search for extraterrestrial life, for which Europa has been a primary target. One possibility to infer the ocean's properties is electromagnetic sounding. This method makes use of the magnetic fields induced by Jupiter's background field, which can be modelled and compared to magnetic field measurements performed by satellite missions. Here, we present a model in which we implement a local water reservoir between the surface and the ocean and calculate the magnetic fields that are induced in both reservoir and ocean. Here, we consider the electromagnetic coupling interaction that takes place between the two water bodies. We investigate if reservoirs can be detected in magnetic field measurements performed with the Europa Clipper \change{satellite}{spacecraft}\remove{, which will enter the Jupiter system in 2030}, \change{and find that reservoirs cannot be detected with Europa Clipper's magnetometer measurements. However, a reservoir can be resolved through the deployment of magnetometers on Europa's surface, where a minimum of two instruments are needed.}{as well as with magnetometers deployed on Europa's surface. The potential detection of liquid water reservoirs could provide insights into the processes inside Europa's icy shell, e.g., the formation of chaos regions visible on Europa's surface.}

%%%%%%%%%%%%%%%%%%%%%%%%%%%%%%%%%%%%%%%%%%%%%%%
%
%  BODY TEXT
%
%%%%%%%%%%%%%%%%%%%%%%%%%%%%%%%%%%%%%%%%%%%%%%%

%%% Suggested section heads:
\section{Introduction} 
The Galilean moon Europa is expected to host a subsurface ocean of liquid salt-water underlying an icy crust. The primary evidence for such an ocean comes from magnetometer measurements obtained during flybys with the Galileo \change{satellite}{spacecraft}. Jupiter's magnetospheric background field was perturbed in Europa's vicinity, which has been explained by an induced magnetic field as a response to the time-varying component of the Jovian background field \cite{kivelson1997europa,khurana1998induced,kivelson2000galileo}. These perturbations require a shallow, conductive layer, which in the context of Europa's geology is interpreted to be a liquid, saline ocean. Gravitational measurements estimated a H$_2$O-shell of $80-170$ km, but could not distinguish between liquid and frozen parts \cite{anderson1998europa}, resulting in a wide range for the estimated icy crust thickness \cite <see e.g.,>[]{turtle2001thickness,schenk2002thickness,lee2005mechanics}.\\
The ocean's properties can be indirectly inferred from magnetic sounding, as they have an influence on its induction response. It is characterized by the induction amplitude $A$ and phase shift $\phi^\mathrm{ph}$, which are governed by the ocean's depth \add{$d$} and thickness \add{$h$}\remove{ $d$ and $h$}, as well as its electrical conductivity $\sigma$. However, this presents a non-unique problem, as different sets of $(d,h,\sigma)$ can yield equal values for $A$ and $\phi^\mathrm{ph}$. Recently, the study of potential ocean worlds with electromagnetic sounding extended to Uranian moons as well \cite{arridge2021electromagnetic, cochrane2021search, weiss2021searching}.\\
Past efforts to model induced magnetic fields on Europa assume the interior structure to be spherically symmetric \cite{zimmer2000subsurface, schilling2007time}. \citeA{styczinski2022perturbation} characterize the electromagnetic induction in an asymmetric ocean, using latitudinal and longitudinal variations due to tidal deformation as a cause for a global asymmetry. Additionally, the icy moon's geological features motivate further local features and depth variabilities. Reservoirs of water-melt could exist within Europa's icy crust, offering a shallow source of liquid water. Such water lenses \change{would}{could} form during the formation of chaos regions \cite{schmidt2011active}, but they also play a role in plume activity, where the refreezing of a reservoir results in an eruption of endogenic material \cite{lesage2020cryomagma}. During the refreezing, the salt concentration within the remaining liquid part of the reservoir increases \cite{lesage2022chemical}. Water vapor plumes \add{may} have been observed in the ultraviolet by the Hubble Space Telescope (HST) near Europa's south pole \cite{roth2014transient, sparks2016probing}. Plasma wave and magnetic field measurements taken during the Galileo E12 and E26 flybys provide additional evidence for plumes on Europa \cite{jia2018evidence,arnold2019magnetic}.  \\
Separated from the ocean, a reservoir generates its own induced field, offering a potential window of detectability with magnetometer measurements. \add{If ocean and reservoir were connected, induced currents could flow between the two bodies. In such cases, our approach does not apply anymore.} Due to the expected small extension of a reservoir, its induced field approaches negligible values at flyby altitudes performed by the Galileo \change{satellite}{spacecraft}. Galileo's magnetometer measurements could thus not resolve such small-scale features. NASA's Europa Clipper \change{satellite}{spacecraft} will enter Jupiter's orbit in 2030 and perform a series of flybys at Europa with altitudes as low as 25 km. In addition, the prospects of a future lander mission would provide an opportunity to measure the magnetic field at the surface.\\
In this work, we calculate the magnetic fields generated by a local reservoir on top of a subsurface ocean and provide an analytical description for the non-linear feedback between both conductive regions, which is introduced in section \ref{Methods}. We determine the magnetic fields that would be measured at 25 km altitude and on Europa's surface\remove{, respectively}. We derive the magnetic field perturbation caused by a reservoir of given size and conductivity by comparing these calculations against the spherically symmetric case and present the results in section \ref{Results}. We end with a summary and discuss our results in the context of the Europa Clipper and JUICE missions, as well as potential future lander missions (section \ref{SaC}).

% Headings should be sentence fragments and do not begin with a
% lowercase letter or number. Examples of good headings are:

\section{Methods}\label{Methods}
In our model, we make use of analytical solutions for the description of electromagnetic induction in spherically symmetric, conductive bodies to model an overall asymmetric geometry of two neighboring bodies; ocean and reservoir. We introduce the fundamental equations in section \ref{Theory} and afterward present in section \ref{sec:coupled} the non-linear coupling mechanism and our analytical approach to solving it.
\subsection{Electromagnetic Induction} \label{Theory}
The Jovian intrinsic field has a tilt of 9.6$^\circ$ relative to Jupiter's rotation axis. It thus appears to be rotating as seen from Europa with the synodic period $T = 11.23$ h. We can express the periodic temporal variation of the inducing field by writing $\vecB_\ind (t) = \vecB_0 e^{-i\omega t}$, where $\vecB_0$ is the field amplitude and $\omega$ its rotation frequency. The physical magnetic field is the real part of $\vecB_\ind(t)$. The induction response to this field is given by the $Q-$response
\begin{linenomath*}
\begin{equation}
    Q = A e^{i\phi^\mathrm{ph}} = \frac{B_\mathrm{i}}{B_\mathrm{e}},
\end{equation}
\end{linenomath*}
where $B_\mathrm{i}$ and $B_\mathrm{e}$ are the induced and inducing complex field coefficients and $A$ and $\phi^\mathrm{ph}$ are the induction amplitude and phase shift. \change{The ratio between the field coefficients is derived by solving Maxwell's equations for a conductivity $\sigma  \neq 0$ and neglecting displacement currents.}{Inside a planetary body within layers of constant conductivity $\sigma$, the magnetic field follows a diffusion equation of the form}
\begin{linenomath*}
\begin{equation} \label{eq:DiffEq}
    \nabla^2 \vecB = \mu \sigma \partial_t \vecB,
\end{equation}
\end{linenomath*}
\add{where $\mu$ is the magnetic permeability. Equation} \ref{eq:DiffEq} \add{can be derived from Maxwell's equations under the assumption that the displacement current is negligible} \cite <e.g.,>{parkinson1983introduction, saur2010induced}. \add{Solving this diffusion equation, with suitable boundary conditions, provides expressions for the ratio between the complex field coefficients. In spherical coordinates, the variations in latitude and longitude may be describved by spherical harmonics of degree $l$ and order $m$}\remove{The spherical variabilities are described in spherical harmonics}, whereas the equation for the radial dependence of the magnetic field is solved by Bessel functions $J_\nu(z)$ of first kind and order $\nu$ with complex argument $z$. \remove{The specific form of $B_\mathrm{i}/B_\mathrm{e}$ depends on the boundary conditions of the considered geometry.} For a sphere with homogeneous conductivity $\sigma$ and radius $r_\res$ we get
\begin{linenomath*}
\begin{equation}
    \frac{B_\mathrm{i}}{B_\mathrm{e}} = -\frac{l}{l+1}\frac{J_{l+3/2}(r_\res k)}{J_{l-1/2}(r_\res k)},
\end{equation}
\end{linenomath*}
while for a conductive shell with outer radius $r_0$ and inner radius $r_1$ (see \citeA{zimmer2000subsurface} for degree $l=1$ and \citeA{saur2010induced} for arbitrary $l$)
\begin{linenomath*}
\begin{equation}
    \frac{B_\mathrm{i}}{B_\mathrm{e}} = -\frac{l}{l+1}\frac{\xi J_{l+3/2}(r_0 k) - J_{-l-3/2}(r_0k)}{\xi J_{l-1/2}(r_0 k) - J_{-l+1/2}(r_0k)},
\end{equation} 
\end{linenomath*}
with
\begin{linenomath*}
\begin{equation}
    \xi = \frac{r_1 k J_{-l-3/2}(r_1 k)}{(2l+1)J_{l+1/2}(r_1 k) - J_{l-1/2}(r_1 k)},
\end{equation}
\end{linenomath*}
where $k$ is the complex wave number with $k^2 = i\mu_0 \omega \sigma$. From this ratio follows $A = \mathrm{abs} (B_\mathrm{i}/B_\mathrm{e})$ and $\phi^\mathrm{ph} = \mathrm{arg} (B_i/B_e)$ with respective ranges $0\leq A\leq l/(l+1)$ and $-\pi/2 \leq \phi^\mathrm{ph} \leq 0$. \add{Here, we note that the definition of the wave vector in} \citeA{parkinson1983introduction}\add{, $k^2=-i\omega\mu\sigma$, results in the modified Bessel equation, which are solved for the modified Bessel functions. The subsequent use of the standard Bessel equations in} \citeA{parkinson1983introduction} \add{yields a positive phase delay, which differs from the negative range in this work.} In the literature, the factor $l/(l+1)$ is occasionally omitted when calculating the induction amplitude, in which case it ranges from 0 to 1. For a perfect conductor $\sigma \rightarrow \infty$, $B_i/B_e$ approaches $l/(l+1)$, thus obtaining $A=1$ and $\phi^\mathrm{ph} = 0$. \add{For a homogeneous sphere with low conductivity and/or small radius, the induction amplitude can be Taylor-approximated around small values for the argument of the Bessel function $rk \rightarrow 0$. This yields the following expression}
\begin{linenomath*}
\begin{equation} \label{eq:Ares}
    A_\res \approx \frac{r_\res^2 \omega \sigma \mu_0}{15},
\end{equation} 
\add{from which we can directly infer the importance of the size of a spherical reservoir when discussing its induction response and thus detectability. The induction amplitude increases with the square of the radius and linearly with both conductivity and rotation frequency of the inducing field.}
\end{linenomath*}
\begin{figure}[ht!]
     \centering
     \includegraphics[width=0.9\textwidth]{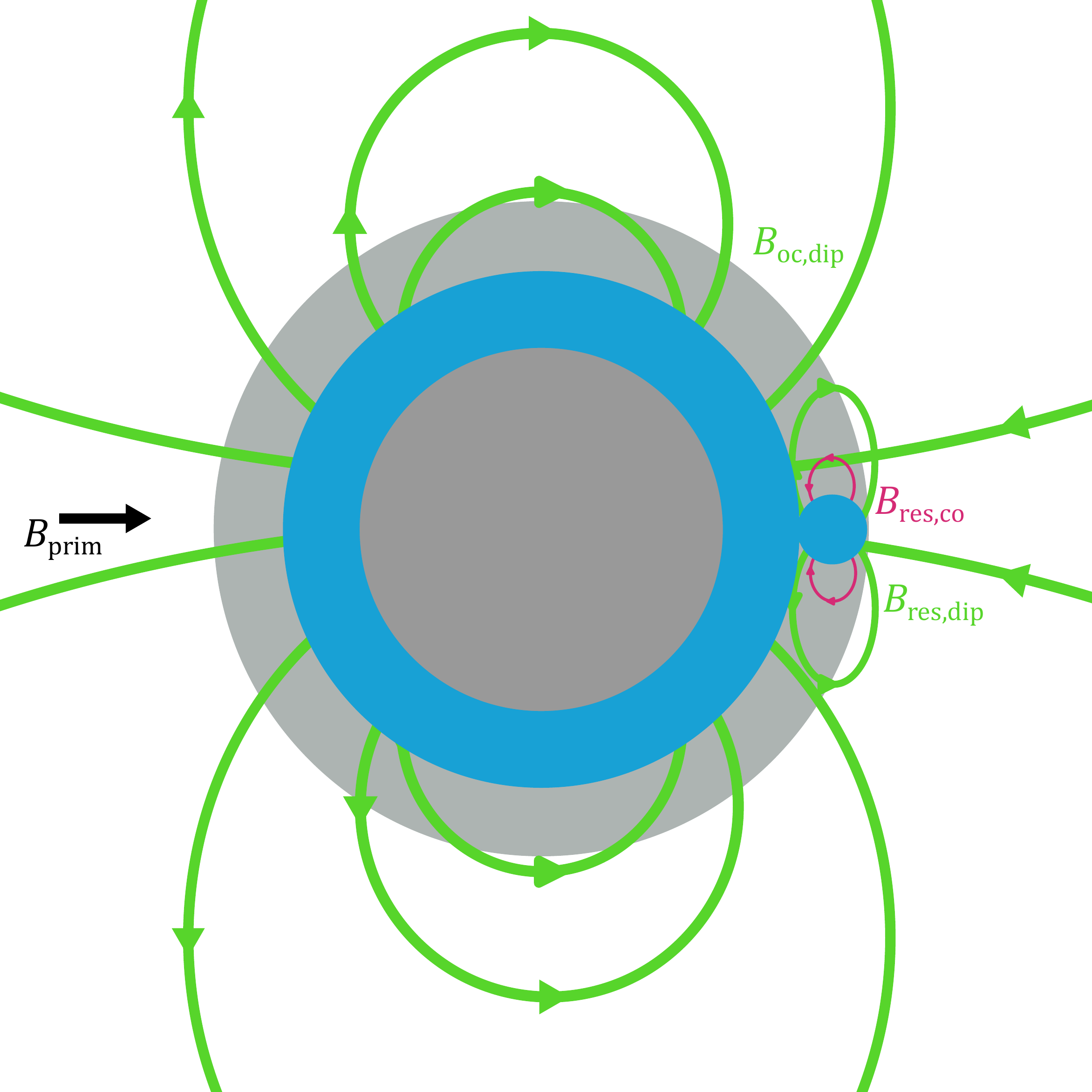}
     \caption{Sketch of the ocean-reservoir model. The Jovian background field $\vecB_\mathrm{prim}$ induces dipoles (green) in the ocean ($\vecB_{\oc,\mathrm{dip}}$) and reservoir ($\vecB_{\res,\mathrm{dip}}$), which induce further magnetic fields (magenta, see e.g., $\vecB_{\res,\mathrm{co}}$). These also act as inducing fields (additional coupled fields have not been drawn for clarity).}
     \label{fig:model}
\end{figure}
\subsection{Coupled Induction} \label{sec:coupled}
The total induced field of two neighboring conductive bodies cannot be described by the sum of their induced dipoles. The induced dipole of Body A induces a magnetic field in Body B and vice versa. This resulting induction response in Body B then acts as an inducing field on Body A, causing a coupling feedback between the two bodies. Due to the proximity of these two bodies, these inducing fields are not homogeneous and thus induce higher order multipole moments. The equations derived in section \ref{Theory} assume spherical symmetry, which holds true for the ocean and reservoir if viewed separately (not the whole system). Application of spherical symmetry to both bodies requires the consideration of two coordinate systems; one with origin in Europa's center (central point of the ocean shell) and the center of an assumed spherical reservoir, respectively. Outside the reservoir and ocean\add{, $\nabla \times \vecB = 0$, assuming the conductivity of the surrounding ice and mantle to be zero. Thus,} the induced magnetic field can be written as a potential field
\begin{linenomath*}
\begin{equation}
    \Phi(r,\theta, \phi,t) = a \sum_{l=1}^{\infty} \sum_{m=0}^l \left(\frac{a}{r} \right)^{l+1} P_l^m(\cos \theta) \left[g_l^m \cos m\phi + h_l^m \sin m\phi  \right] e^{-i\omega t},
\end{equation}
\end{linenomath*}
\add{in Europa IAU coordinates $(r,\theta,\phi)$}, where $g_l^m$ and $h_l^m$ are the internal Gauss coefficients of degree $l$ and order $m$. We also express the inducing field as a potential
\begin{linenomath*}
\begin{equation}
    \Psi(r,\theta, \phi,t) = a \sum_{l=1}^{\infty} \sum_{m=0}^l \left(\frac{r}{a} \right)^l P_l^m(\cos \theta) \left[q_l^m \cos m\phi + s_l^m \sin m\phi  \right] e^{-i\omega t}.
\end{equation}
\end{linenomath*}
For both potentials, the physical magnetic field is described by the real part of the complex description. Here, $q_l^m$ and $s_l^m$ denote the external Gauss coefficients of degree $l$ and order $m$, which are obtained by integration 
\begin{linenomath*}
\begin{equation}\label{eq:gext}
    \begin{Bmatrix} 
    q_l^m \\
    s_l^m
    \end{Bmatrix} = -\frac{2l+1}{4\pi l}\int_0^{\pi} \int_0^{2\pi} d\phi d\theta \sin \theta P_l^m(\cos \theta) B_r(\theta,\phi) \begin{Bmatrix}
        \cos m\phi\\
        \sin m\phi
    \end{Bmatrix},
\end{equation}
\end{linenomath*}
where $B_r(\theta,\phi)$ is the radial component of the inducing field of on the surface of the conducting body in a coordinate system in the center of that body. This means that we need to transform the ocean's induction response in reservoir-centered coordinates and calculate the radial component to obtain the external coefficients of that inducing field. This process is done vice versa for the ocean's induction response to the reservoir. A detailed description of the coordinate transformation can be found in \ref{app:Transform}. \change{Omitting the radial dependence and considering each degree $l$ of the potential description separately, we can multiply the inducing field potential with the $Q$-response}{The $Q-$response can also be written as the ratio between complex Gauss coefficients}
\begin{linenomath*}
\begin{equation}
    Q_l = \frac{\iota_l^m}{\epsilon_l^m},
\end{equation}
\end{linenomath*}
\add{where $\iota_l^m$ are the internal and $\epsilon_l^m$ the external complex coefficients} \cite{constable2004observing, saur2010induced}. \add{This ratio can also be derived by taking the ratio between induced and inducing field potentials at the surface in their complex description. Thus, to obtain the induced field, we can mulitply the $Q-$response with the inducing field at the surface. As the $Q-$response is a function of degree $l$, this multiplication is done for each degree of the multipole potential seperately}
\begin{linenomath*}
\begin{equation}
 \begin{split} \label{eq:QPsi}
     \Phi_l =& Q_l \Psi_l = A_l e^{i\phi_l}  \sum_{m=0}^l P_l^m(\cos \theta) \left[q_l^m \cos m\phi + s_l^m \sin m\phi  \right] e^{-i\omega t} \\
     =&  \sum_{m=0}^l P_l^m(\cos \theta) \left[A_l q_l^m \cos m\phi + A_l s_l^m \sin m\phi  \right] e^{-i(\omega t - \phi^\mathrm{ph}_l)},
 \end{split}
\end{equation} 
\end{linenomath*}
which reveals crucial properties of the induced field:
\begin{enumerate}
    \item Its internal coefficients are related to the external coefficients of the inducing field via 
\begin{linenomath*}
\begin{equation} \label{eq:Gausscoefficients}
    \begin{Bmatrix} 
    g_l^m \\
    h_l^m
    \end{Bmatrix} = A_l \begin{Bmatrix} 
    q_l^m \\
    s_l^m
    \end{Bmatrix},
\end{equation}
\end{linenomath*}
i.e., under assumption of spherical symmetry, an inducing field of degree $l$ and order $m$ only induces a field of same degree and order.
    \item The induction response is subject to \add{a} temporal \change{retardation}{delay}, i.e., the induced field at time $t$ is the response to the inducing field at an earlier time $t+\phi^\mathrm{ph}_l/\omega$.
    \item The $Q$-response changes with degree $l$, i.e., each multipole moment of the induction response at time $t$ is induced at a different time $t+\phi^\mathrm{ph}_l/\omega$.
\end{enumerate}
In the following, we will thoroughly describe the first few iterations of the coupling feedback, starting with the induced dipole. We introduce the iteration step $n$, with which we denote the Gauss coefficients and induced magnetic fields for clarity. $n=1$ corresponds to the induced dipoles with internal coefficients $(g_1^1)^{(1)}, (h_1^1)^{(1)}$, $n=2$ are the fields induced by the dipoles, $n=3$ is the induction response to the induced fields of $n=2$ and so on.

\subsubsection{Calculating the Induced Dipole, $n=1$} \label{dipole}
For the induced dipoles, the series only contains $l=1$ terms, as the Jovian background field can be assumed to be homogeneous in Europa's vicinity, i.e., spatial variabilities can be omitted. We assume that the background field is elliptically polarized in the \change{$x-y$-plane}{$xy$-plane}\remove{, where $x$ points in the direction of corotational flow and $y$ towards Jupiter}, thus writing $\vecB_\Jup(t) = (B_{\Jup,x}(t), B_{\Jup, y}(t), 0)$ \cite <see>[for more information]{seufert2011multi}. Further specifications of the background field follow in section \ref{sec:inducingfield}. Due to the homogeneity, the external coefficients of the inducing field are a dipole with $q_1^1(t+\phi^\mathrm{ph}_1/\omega) = q_\Jup(t+\phi^\mathrm{ph}_1/\omega) = - B_{\Jup,x}(t+\phi^\mathrm{ph}_1/\omega)$ and $s_1^1(t+\phi^\mathrm{ph}_1/\omega) = s_\Jup(t+\phi^\mathrm{ph}_1/\omega) = - B_{\Jup,y}(t+\phi^\mathrm{ph}_1/\omega)$, where the temporal retardation is also taken into account. In the equation above and throughout this work, we will include the time-dependency of the potentials in the external and internal Gauss coefficients, so that $g_l^m(t) = g_l^m e^{-i\omega t}$. As such, from equation \eqref{eq:QPsi}, one can infer that the induced field governed by internal coefficients $(g_1^1)^{(1)} (t),(h_1^1)^{(1)} (t)$ acts as the response to the inducing field at an earlier time $q_\Jup (t+\phi^\mathrm{ph}_1/\omega), s_\Jup (t+\phi^\mathrm{ph}_1/\omega)$, where the phase shift $\phi^\mathrm{ph}$ is negative within the notation of this work. Rewriting equation \eqref{eq:Gausscoefficients} to account for the temporal variability yields the internal coefficients of the induced dipole at time $t$ 
\begin{linenomath*}
\begin{equation} \label{eq:Gauss}
    \begin{Bmatrix}
        (g_1^1)^{(1)} e^{-i\omega t} \\
        (h_1^1)^{(1)} e^{-i\omega t}
    \end{Bmatrix} = A_1 \begin{Bmatrix}
        q_\Jup e^{-i(\omega t+\phi^\mathrm{ph}_1)} \\
        s_\Jup e^{-i(\omega t + \phi^\mathrm{ph}_1)}
    \end{Bmatrix}.
\end{equation}
\end{linenomath*}

\subsubsection{Calculating Induced Fields from Coupling, $n>1$}
The coupled induction responses contain higher degree terms due to the inhomogeneity of the inducing field. After performing the coordinate transformation, the ocean's dipole is no longer a dipole field on the surface of the \change{other body}{reservoir}, but generates a full set of multipole moments $q_l^m, s_l^m$. \add{These drive induced fields from the reservoir that further act as an inducing field on the ocean. Solving for the overall induction response requires an iterative approach.} To distinguish between ocean (oc) and reservoir (res), we further add subscripts to the Gauss coefficients attributed to the respective body. For each degree, we need to consider the coupling feedback separately as the phase shift $\phi^\mathrm{ph}_{\res,l}$ is a function of degree, meaning that the 'starting point', i.e., the time at which we consider the Jovian background field, changes for the different multipole contributions of the coupled induced field. For the multipole moment of degree $l$, we calculate the external coefficients of the Jovian background field at time $t+(\phi^\mathrm{ph}_{\oc,1}+\phi^\mathrm{ph}_{\res,l})/\omega$. Here, $\phi^\mathrm{ph}_{\oc,1}$ is the temporal \change{retardation}{delay} due to the induced dipole discussed in section \ref{dipole} and $\phi^\mathrm{ph}_{\res,l}$ is the additional \change{retardation}{delay} due to the coupling. From here, we obtain the internal coefficients of the ocean's dipole $(g_{\oc,1}^1)^{(1)}(t+\phi^\mathrm{ph}_{\res,l}/\omega)$ and $(h_{\oc,1}^1)^{(1)}(t+\phi^\mathrm{ph}_{\res,l}/\omega)$.
We transform the resulting field into reservoir-centered coordinates and obtain the external Gauss coefficients of the inducing field with which we calculate the reservoir's internal Gauss coefficients of the second iteration
\begin{linenomath*}
\begin{equation}
    \begin{Bmatrix} 
    (g_{\res,l}^m)^{(2)} (t) \\
    (h_{\res,l}^m)^{(2)} (t)
    \end{Bmatrix} = A_{\res,l} \begin{Bmatrix} 
    (q_{\oc,l}^m)^{(1)}(t+\phi^\mathrm{ph}_{\res,l}/\omega) \\
    (s_{\oc,l}^m)^{(1)}(t+\phi^\mathrm{ph}_{\res,l}/\omega)
    \end{Bmatrix},
\end{equation} 
\end{linenomath*}
The ocean's induction response of the second iteration is obtained analogously. Iteration step $n=3$ introduces an additional layer of complexity. In the second iteration, we obtained a full set of external coefficients $(q_{\oc,l}^m)^{(1)}, (s_{\oc,l}^m)^{(1)}$ with $l=[1,2,3,...,l_\mathrm{max}]$ due to the transformation of the ocean dipole onto the surface of the reservoir, where $l_\mathrm{max}$ is the maximum degree considered in our study. From this, we obtain an induced field that contains various multipole moments $(g_{\res,l}^m)^{(2)}, (h_{\res,l}^m)^{(2)}$, where $l = [1,2,3,...,l_\mathrm{max}]$. $l_\mathrm{max}$ is formally required to approach infinity. The choice of $l_\mathrm{max}$ to obtain a solution with a prescribed precision is discussed in section \ref{sec:lmax}. To calculate the induction response to this induced field in the third iteration, we need to transform each multipole degree $l$ individually, as the phase shift is a function of $l$. Each multipole degree $l$ gives rise to a full set of multipole moments $(\Tilde{q}_{\res,l'}^m)^{(2)}(l), (\Tilde{s}_{\res,l'}^m)^{(2)}(l)$ after performing the coordinate transformation, with $l' = [1,2,3,...,l_\mathrm{max}]$. The sum of all contributions yields the external coefficients that fully describe the inducing field at a given time.
\begin{linenomath*}
\begin{equation}
 \begin{split}
     \sum_{l'=1} (\Tilde{q}_{l}^m)^{(2)}(l') = & (q_{l}^m)^{(2)} \\
     \sum_{l'=1} (\Tilde{s}_{l}^m)^{(2)}(l') =& (s_{l}^m)^{(2)}  .
 \end{split}
\end{equation} 
\end{linenomath*}
With each iteration, the induced field is subject to further \change{retardation}{delay} and the strength of the induction response decreases due to two factors; multiplication with an induction amplitude $A<1$ and decrease of the inducing field strength due to the distance between the two bodies. This effect ensures that with increasing iteration steps the solution converges to the full solution, where mutual induction effects are consistently described. For perfectly conducting bodies, $A=1$, which results in a 'slower' convergence, i.e., the system requires more coupling iteration to reach its equilibrium state than in the finite case.

\subsubsection{Mauersberger-Lowes Spectrum}\label{sec:Mauersberger}
To control the iteration steps necessary to describe mutual induction responses, we consider the Mauersberger-Lowes spectrum, where the magnetic power per degree and iteration step $(R_l)^{(n)}$ is defined as \cite{langel1982geomagnetic}
\begin{linenomath*}
\begin{equation}\label{eq:Mauersberger}
    (R_l)^{(n)} = (l+1) \sum_{m=0}^l \left[((g_l^m)^{(n)})^2 + ((h_l^m)^{(n)})^2 \right].
\end{equation}
\end{linenomath*}
Iteration steps that contribute insignificantly to the total induction response of the system will have a magnetic power that is multiple orders of magnitude below that of the first iteration step; $(R_l)^{(n)} << (R_1)^1$. \add{Multipole moments with Gauss coefficients on the order of $10^{-2}$ nT will result in very weak magnetic fields that cannot be resolved. For a dipole moment, this correlates to a magnetic power on the order of $10^{-4}$ nT$^2$. Thus, multipole degrees or entire iterations that lie below this value can be neglected.}

\subsubsection{Numerical Precision and Choice of $l_\mathrm{max}$} \label{sec:lmax}
In the numerical implementation of our model, the multipole degrees cannot be calculated to an infinite degree. Thus, after transforming the ocean's induction response into reservoir-centered coordinates\add{, $B_{r,\mathrm{transformed}}(\theta,\phi)$}, to obtain the multipole moments of the inducing field, this field is approximated by a finite set of coefficients $q_l^m,s_l^m$\add{, with $B_{r,\mathrm{gauss}}(\theta,\phi)$}. To obtain a sufficiently accurate solution, we calculate the inducing field from Gauss coefficients of degree up to $l_\mathrm{max}$ across the surface and compare the result to the exact description from the coordinate transformation. We determine the root-mean-square (RMS) 
\begin{linenomath*}
\begin{equation}
    RMS = \sqrt{\frac{\sum_{i,j}(B_{r,\mathrm{gauss}}(\theta_i,\phi_j)-B_{r,\mathrm{transformed}}(\theta_i,\phi_j))^2}{N_{ij}}},
\end{equation}
\end{linenomath*}
\add{where $N_{ij}$ is the number of gridpoints,} and employ a limit of \change{$10^{-5}$}{$10^{-2}$} nT, \change{This choice arises from the consideration of weak induction signatures for the reservoir, which can be}{as magnetic fields below that value will not be detectable with magnetometer measurements. However, we still calculate the induced fields to the 5th significant figure, as the considered induced fields can be} on the order of $10^{-4}$ nT for the low conductivity limit (see section \ref{sec:25km}).

\subsection{Model Implementation and Parameter Space} \label{sec:pspace}
 \subsubsection{Inducing Background Field} \label{sec:inducingfield}
 We approximate the inducing part of the Jovian background field \add{$\vecB_\mathrm{J}(t)$} to be elliptically polarized \cite <see Figure 1 in>[]{khurana1998induced} with
 \begin{align}\label{eq:Bfield}
     B_{\mathrm{J},x}(t) &= B_{0,x} \cos(\omega t + \phi^\mathrm{ph}_x) \nonumber \\
     B_{\mathrm{J},y}(t) &= B_{0,y} \cos(\omega t + \phi^\mathrm{ph}_y) \\
     B_{\mathrm{J},z}(t) &= 0 \nonumber,
 \end{align}
 \add{in Europa IAU coordinates}, where $\vecB_0 = (-217,64,0)$ nT, $\phi^\mathrm{ph}_x = 0$, and $\phi^\mathrm{ph}_y = -\pi/2$. \add{Note that the magnetic field in} \citeA{khurana1998induced} \add{is given in EPhiO coordinates, which correspond approximately to
 $x\rightarrow y$ and $y\rightarrow -x$ in IAU coordinates. For the purpose of this paper, we neglect the small misalignment of unit vector directions in EPhiO and Europa IAU coordinates.}\remove{The physical magnetic field is
 the real part of equation} We will choose \change{$\omega t_\mathrm{obs} = \pi/2$}{$\omega t_\mathrm{obs} = \pi$} in section
 \ref{Results}, where we show the results of our coupled model. This choice leads to the maximum
 background field for our induction studies.
 \subsubsection{Geometric Parameters} \label{sec:Geometry}
 As chaos terrain is mostly found near equatorial regions \cite{greenberg1999chaos}, we assume a reservoir position in the \change{$x-y$-plane}{$xy$-plane}. We choose a position parallel to the \change{$y$}{$x$}-axis, which also aligns with the magnetic background field at the chosen time. As the main focus of this work lies on the induction signature of a reservoir of different sizes and conductivities, \change{we keep the parameters of the ocean fairly unchanged in our study}{the ocean's induction amplitude remains constant to ensure similar induction responses throughout the study.} We assume a fixed sea floor depth of 150 km, corresponding to an inner radius $r_1 = 1410$ km. The outer radius will be changed accordingly to the radius of the reservoir so that the reservoir spans across the entire icy crust, \add{but still does not overlap with the global ocean. In this way, our study gives an upper limit to the expected small signals. The only exceptions to this geometry are Figures} \ref{fig:SC_Vectorfield} and \ref{fig:nonSC_Vectorfield}, where a small gap is added to highlight additional effects of the ocean-reservoir system. \add{As changing the outer radius effects the ocean's induction amplitude, its conductivity must be adjusted accordingly (see section} \ref{sec:EC}). \add{It should be noted that while this method keeps the induction amplitude constant, this does not hold true for the phase shift.} We will consider reservoir radii ranging from $5-20$ km, accordingly to crust thicknesses found in literature.
 \subsubsection{Ice shell thickness} \label{sec:iceshell}
 The thickness of Europa's icy crust is a crucial parameter in the description of its induction response, as the ratio $(r_0/r_\mathrm{m})^3$ governing the induced dipole field is larger for a shallow ocean compared to a deep one, resulting in a stronger induced magnetic field around Europa. Different methods have been used to infer the crust's thickness, yielding varying estimates. A range of those estimates is summarized in Table \ref{tab:IceThickness}. It is noteworthy that, as a consequence of the non-uniqueness of the induction problem and unknown conductivity, the bounds found with the induction method are less constraining than other methods. 
 \begin{table}[hb!t]
  \caption{A collection of literature values for the ice shell thickness $h$.}
  \centering  
  \begin{tabular}{c c c}
   $h$ /km & Reference & Method \\ \hline \hline
   $\leq 15$ & \citeA{hand2007empirical} & Induction \\
   $\leq 100$ & \citeA{zimmer2000subsurface} & Induction \\
   $\geq 4$ & \citeA{turtle2001thickness} & Impact crater modelling \\
   $5-10$ & \citeA{silber2017impact} & Impact crater modelling \\
   $>19$ & \citeA{schenk2002thickness} & Impact crater observation \\
   %$\approx 10 $ & Moore et al., 2001 & Impact crater observation \\
   $20-25$ & \citeA{tobie2003tidally} & Tidal dissipation model \\
   $\leq 3$ & \citeA{lee2005mechanics} & Cycloid crack formation  \\
   $\approx 25$ & \citeA{prockter2000folds} & Crustal cycling
  \end{tabular}
  \label{tab:IceThickness} 
 \end{table}
 \subsubsection{Electrical Conductivity} \label{sec:EC}
 The electrical conductivity driven by dissolved ions can generally be expressed as a function of pressure, temperature, and concentration of the respective salt \cite{mccleskey2012comparison,pan2020electrical, pan2021electrical}. A key issue about Europa's ocean is the uncertainty about its chemical composition \cite{mckinnon2003sulfate, ligier2016vlt, trumbo2019sodium,trumbo2022new}. As recent UV observations favor an ocean composition rich in sodium chloride, we focus on the conductivity range achieved with it. \citeA{hand2007empirical} inferred conductivities of up to 30 S/m. It should, however, be noted that they used a fit for sea salt, of which the main contributor is sodium chloride at approximately 90\%.\remove{ provide a numerical model for the electrical conductivity based on laboratory measurements, with which they provide conductivities over the high pressure-low temperature range, yielding conductivities of up to 13 S/m within temperatures of $255-290$ K and pressures of up to 1000 MPa.} For our parameter study, values ranging from $0.5-30$ S/m will be considered for the reservoir, covering a low conductivity limit as well as the conductivity at saturation. As we vary the ocean's outer radius, keeping a fixed ocean conductivity $\sigma_\oc$ throughout all simulations would result in different induction amplitudes for the ocean and thus influence the reservoir's induction differently between individual runs. Thus, we adapt the conductivity of the ocean so that the induction amplitude remains constant at $A_\oc=0.91$. This \add{is done by computing the induction amplitude prior the study and adjusting the conductivity until $A_\oc$ reaches 0.91,} \change{value corresponds}{corresponding} to a conductivity of $\sigma_\oc \approx 0.5 \pm 0.1$ S/m for ocean thicknesses in the range of $110-140$ km, which has been derived in \citeA{schilling2007time}. Other values for the ocean's conductivity are plausible, as is evident by the broad conductivity range found in literature cited within this section. This is due to the non-uniqueness of the problem, where different choices for the ocean's thickness, depth, and conductivity can result in the same induction amplitude.
 
 \subsubsection{Frequency of the Inducing Field}
\begin{figure}
    \centering
    \includegraphics[width=0.9\textwidth]{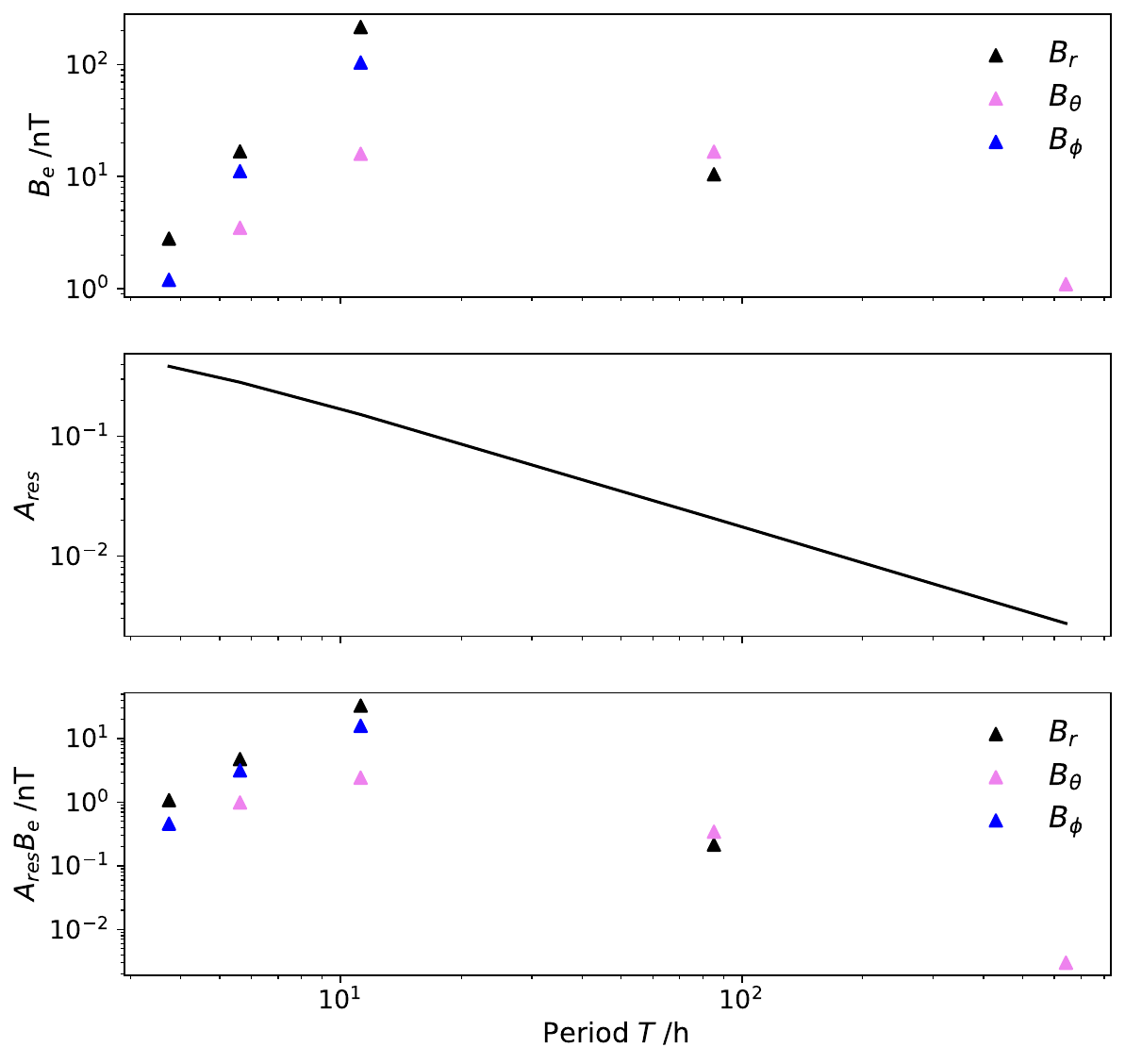}
    \caption{Magnetic field amplitudes of the periodic signals in \change{spherical components}{System III coordinates} (top). Induction response of a 20 km-diameter reservoir ($\sigma_{res} = 10$ S/m) as a function of period (middle). Resulting induction responses \add{of the reservoir} to the magnetic field amplitudes at their respective periods (bottom).}
    \label{fig:Multifrequency}
\end{figure}

In addition to Jupiter's synodic rotation period, Europa is subject to further periodicities with various magnetic field amplitudes, which allow for electromagnetic sounding at multiple frequencies \cite{seufert2011multi}. This is of particular interest for the detection of a shallow reservoir, as its induction amplitude increases with frequency, e.g., higher order harmonics of Jupiter's synodic period at $5.62$ h and $3.33$ h, respectively. However, the amplitudes of the inducing field at those frequencies are one to two orders of magnitudes lower compared to the amplitude at the $11.23$ h period, resulting in an overall weaker induction response at these frequencies. This is shown in Figure \ref{fig:Multifrequency}, where the reservoir's induction response in spherical \change{components}{coordinates} is given for various periodicities. In addition to the synodic periods, we also present the induction signals at $85.22$ h and $641.90$ h, corresponding to Europa's orbital period and the solar rotation period. The induced fields resulting from these signals are weaker. Thus, we will only consider the magnetic fields induced by the synodic period $T = 11.23$ h throughout calculations in this work.
 
\section{Results} \label{Results}
To highlight the effects of the coupled induction, we present the physics \add{in the illustrative case} of a perfectly conducting ocean and reservoir, before showing the effects of coupled induction in finite cases. Afterward, we investigate the detectability of reservoirs with radii and conductivities discussed in sections \ref{sec:Geometry} and \ref{sec:EC} at 25 km altitude and at the surface.
\subsection{Physics of the Coupled System}
\begin{figure}
    \centering
    \includegraphics[width=\textwidth]{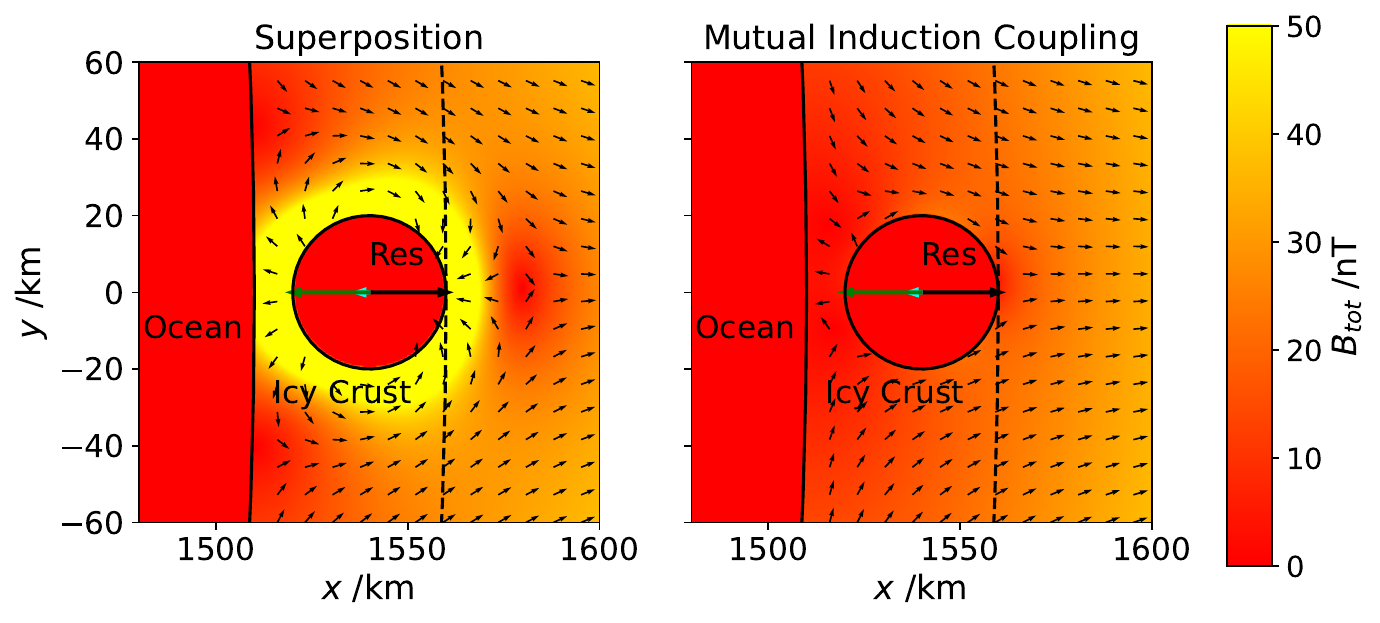}
    \caption{Magnetic field \change{lines}{components in the $xy$-plane} in the ocean-reservoir environment with radius $r_\res = 20$ km, assuming both ocean and reservoir are perfectly conducting. The left panel shows the \change{field lines}{vector field} only after the first iteration, i.e., a superposition of the induced dipole fields in the ocean and reservoir, whereas the right panel visualizes the field \remove{lines} after coupling iteration $n=2$. \add{The arrows representing the magnetic field are normalized in length and do not represent the strength of the magnetic field, which is instead color coded in the background. The arrows in the center indicate the orientations of the background field (black), ocean induced field (green), and reservoir induced field (cyan), where the length indicates the strength relative to the background field. The dashed black line visualizes the surface boundary.}}
    \label{fig:SC_Vectorfield}
\end{figure}

\begin{figure}
    \centering
    \includegraphics[width=\textwidth]{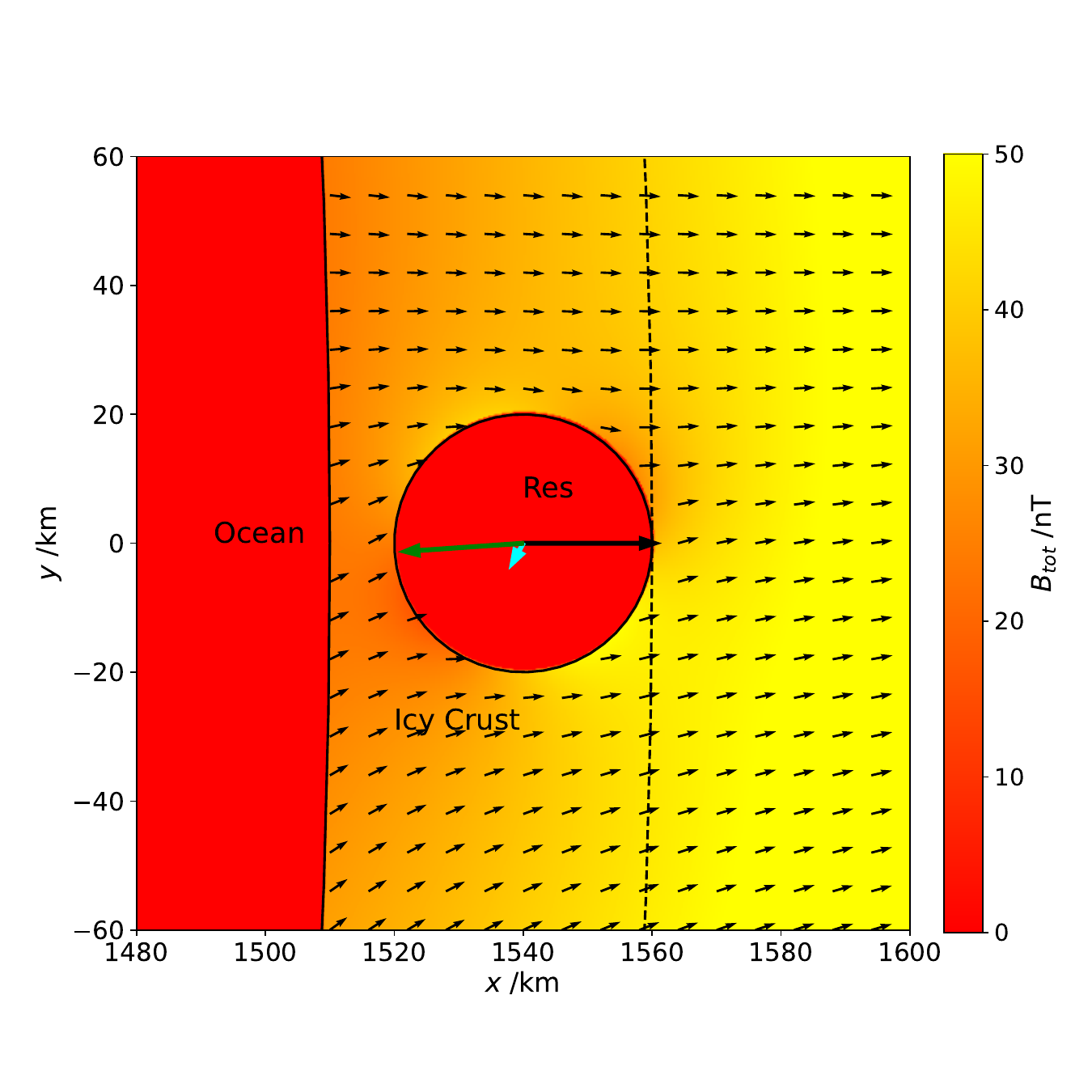}
    \caption{Magnetic field \change{lines}{components in the $xy$-plane} in the ocean-reservoir environment after mutual induction coupling for a reservoir with radius $r_\res = 20$ km and finite conductivity $\sigma_\res = 30$ S/m, \add{where the induction response of the system follows equation} \ref{eq:InductionResponse}. \add{The arrows representing the magnetic field are normalized in length and do not represent the strength of the magnetic field, which is instead color coded in the background. The arrows in the center indicate the orientations of the background field (black), ocean induced field (green), and reservoir induced field (cyan), where the length indicates the strength relative to the background field. The dashed black line visualizes the surface boundary.}}
    \label{fig:nonSC_Vectorfield}
\end{figure}

Figure \ref{fig:SC_Vectorfield} visualizes the magnetic field \change{lines}{components in the $xy$-plane} in the vicinity of a perfectly conducting reservoir, with and without coupling effects. It is apparent that only with consideration of coupled induction, the reservoir behaves like a perfectly conducting body, i.e., \change{magnetic field lines}{the inducing field} \change{do}{does} not penetrate the body. We added a small gap between the ocean and reservoir in the example shown in Figure \ref{fig:SC_Vectorfield}. This is to highlight that in addition to not permitting the magnetic field to penetrate the conducting bodies, due to $\sigma \rightarrow \infty$, the area between reservoir and ocean is also nearly completely shielded from the magnetic field at a close distance.\\
In a realistic scenario, the ocean's and reservoir's induction amplitudes are smaller than one. In addition to this, their phase shift is non-zero, resulting in induction responses that are no longer antiparallel to the inducing field, which is shown in Figure \ref{fig:nonSC_Vectorfield}. Here, the total magnetic field is stronger than in the perfectly conducting case due to the weaker induction response and the \change{field lines}{$xy$-components} are not aligned along the $y$-axis as a result of the phase shift.  
\subsection{Coupling Interaction Strength}
\begin{figure}
    \centering
    \includegraphics[width=0.9\textwidth]{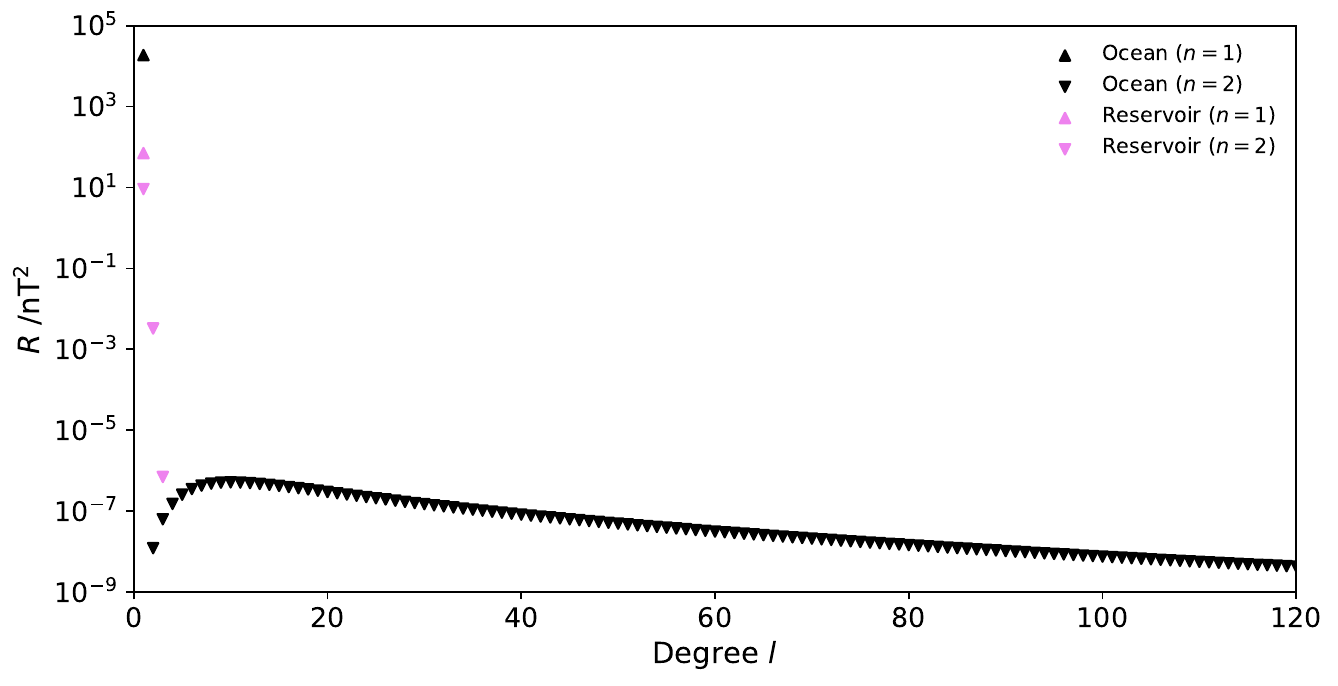}
    \caption{Mauersberger-Lowes spectrum for the ocean (black) and reservoir (violet) for multipole degrees up to $l_\mathrm{max}=120$. Upward triangles show the dipoles induced by the Jovian background field. Downward triangles represent the induced fields of second iteration.}
    \label{fig:Mauersberger}
\end{figure}

To obtain a sufficiently accurate and numerically efficient description of the coupling processes, it is important to look at the contributions of each coupling iteration. Figure \ref{fig:Mauersberger} shows the Mauersberger-Lowes spectrum calculated with equation \ref{eq:Mauersberger} for both the ocean and reservoir induced fields up to the first coupled field (iteration step $n=2$). For the ocean, the main contribution is generated by the dipole response to the Jovian background field \add{with a magnetic power of approximately $10^4$ nT$^2$}. Due to the geometry, the reservoir induces higher degrees in the ocean, where the power reaches a maximum of approximately $10^{-6}$ nT$^2$ at degree $l = 7$ and decreases thereafter. Compared to the dipole term \add{of first iteration $n=1$} with $10^4$ nT\add{$^2$}, the coupling from the ocean to the reservoir\add{, i.e., the induced field of second iteration $n=2$,} is negligible, \add{with a radial component in the order of $10^{-2}$ nT at the surface of the reservoir. It} is thus excluded in any further calculations and results\add{, as such weak induced fields will not be detectable}. \\
The reservoir's induction response \remove{to the ocean is primarily a dipole, however degrees up to degree $l=4$ generate minor contributions as well due to the inhomogeneous nature of the inducing field and must be considered to meet the prescribed precision of $10^{-5}$ nT introduced in section.}\add{of first iteration $n=1$ is the dipole response to Jupiter's background field and thus only has a degree $l=1$ contribution. The magnetic power of the induced dipole in the reservoir is around two orders of magnitude smaller than the ocean's induction response. This is a direct consequence of the reservoir's small extension and thus induction amplitude, as following from equation} \ref{eq:Ares}. \add{However, the reservoir induced field of second iteration $n=2$ has significant contributions at degrees $l=1$ and $l=2$ due to the inhomogeneous nature of the inducing field from the ocean and must be considered to describe the mutual feedback and to meet the prescribed precision of $10^{-2}$ nT introduced in section} \ref{sec:lmax}. \add{Thus, the reservoir is strongly coupled to the ocean, whereas the influence from the reservoir on the induction response of the ocean is overall negligible (see small $R$ values for $l>1$ for the ocean in Figure} \ref{fig:Mauersberger}). The total induction response of the system can be described as a sum of three induction responses
\begin{linenomath*}
\begin{equation}\label{eq:InductionResponse}
    \vecB_\mathrm{ind, tot} = \vecB_\oc^{(1)} + \vecB_\res^{(1)} + \vecB_\res^{(2)}. 
\end{equation}
\end{linenomath*}
For the geometries considered in this model, the reservoir is coupled to the ocean and reacts considerably to its induced dipole. The coupling of the ocean to the reservoir is negligibly small due to the difference in size.

\subsection{Flyby at 25 km Altitude}\label{sec:25km}
\begin{figure}
    \centering
    \includegraphics[width=0.9\textwidth]{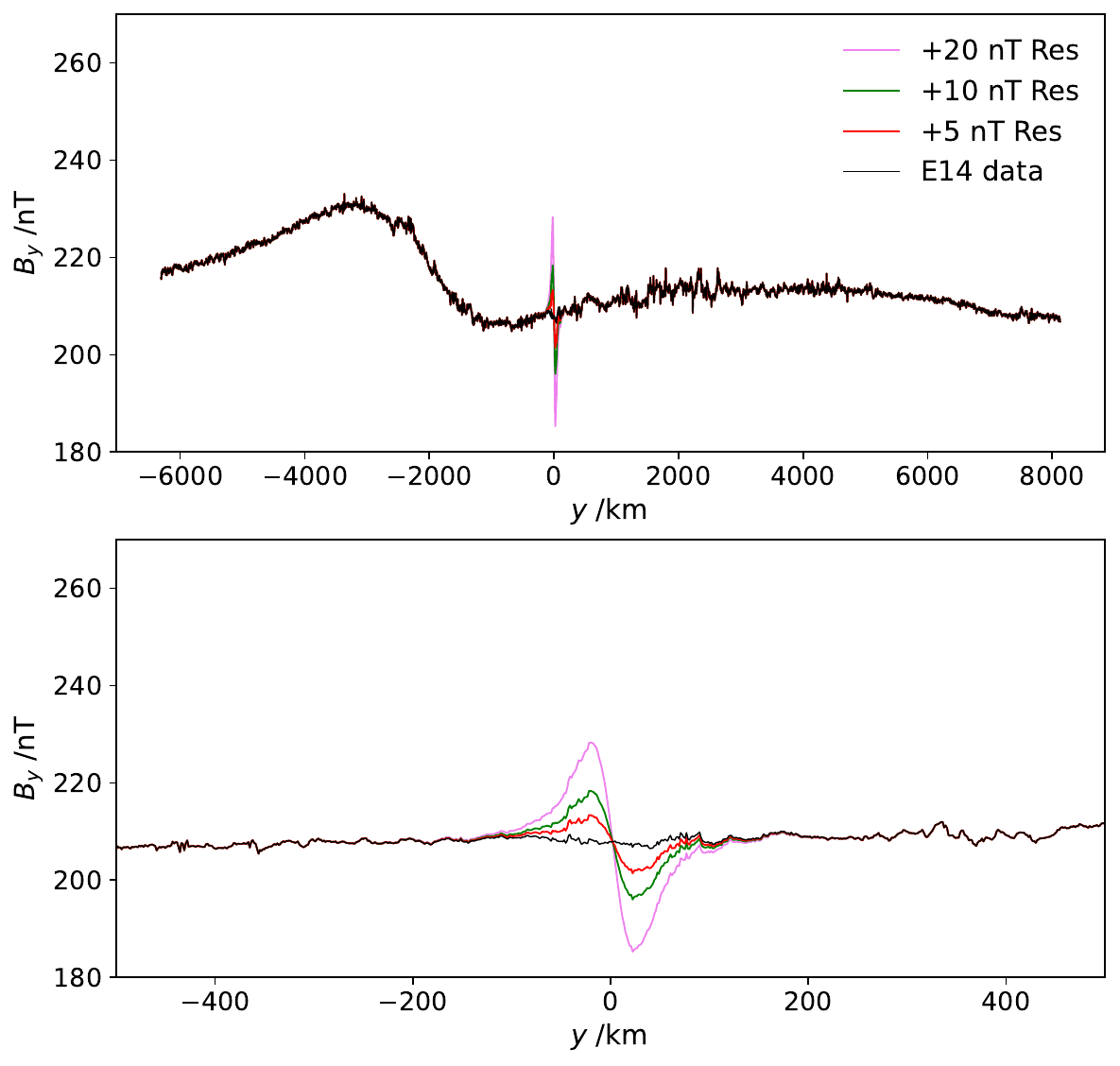}
    \caption{Magnetometer measurements of the $B_y$ component during the Galileo E14 flyby (black). Additionally, artificial signals of a perfectly conducting reservoir with 5 nT (red), 10 nT (green), and 20 nT (violet) amplitude at 25 km altitude have been superimposed to the data to test if these signals are distinguishable from the fluctuations. The bottom panel shows a zoom into the 1000 km around C/A to show the reservoir's induction characteristic.}
    \label{fig:E14}
\end{figure}

\begin{figure}
    \centering
    \includegraphics[width=0.9\textwidth]{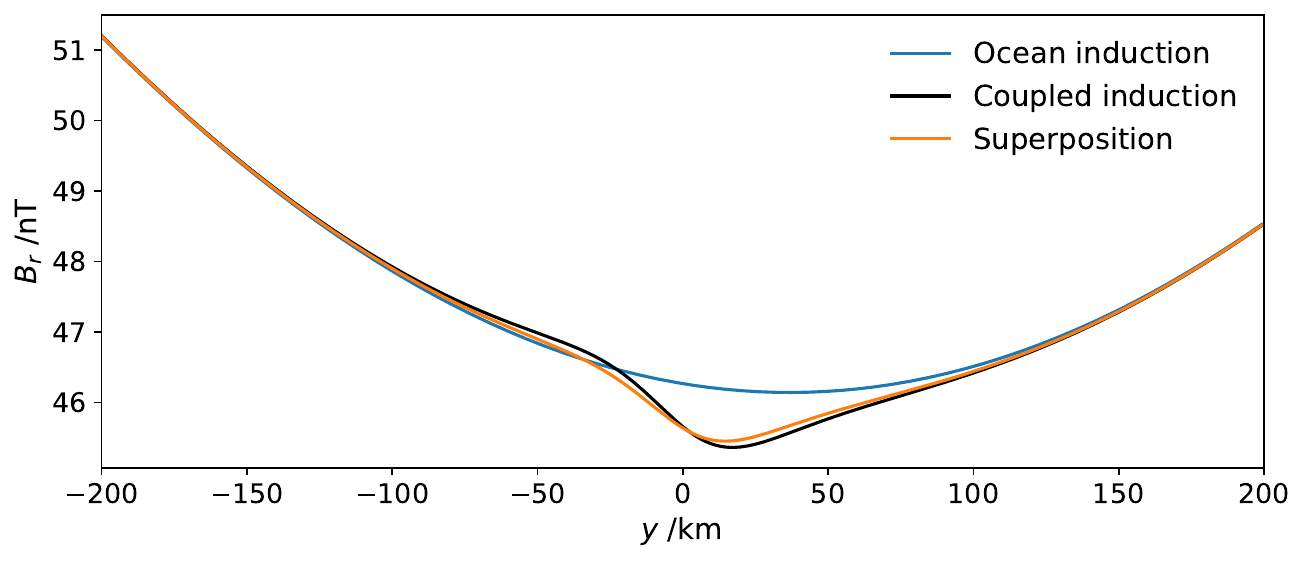}
    \caption{Radial component of the \add{sum of inducing and induced} magnetic field \add{$B_r = B_{\mathrm{ind},r} + B_{\mathrm{J},r}$, where $B_{\mathrm{ind},r}$ follows equation} \ref{eq:InductionResponse} \add{and $B_{\mathrm{J},r}$ is obtained by considering equation} \ref{eq:Bfield} at $\omega t=\pi$, during a hypothetical flyby with 25 km altitude at closest approach with reservoir parameters $\sigma = 30$ S/m and $r_\res = 20$ km.}
    \label{fig:Flyby}
\end{figure}

\begin{figure}
    \centering
    \includegraphics[width=0.8\textwidth]{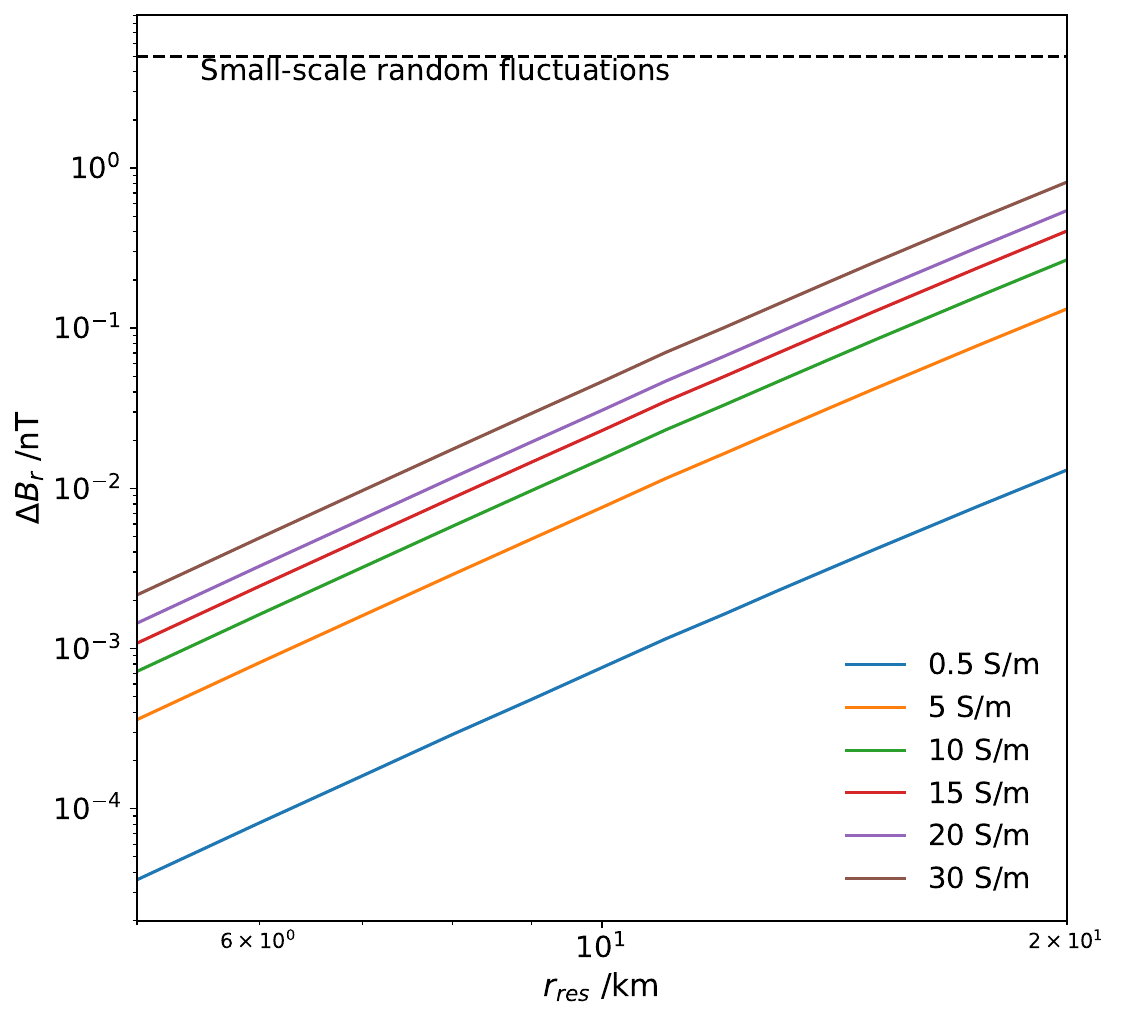}
    \caption{Magnetic induction signature caused by reservoir at 25 km altitude as a function of radius for conductivities ranging from 0.5 to 30 S/m. The dashed horizontal line at 5 nT represents the minimum at which reservoirs could be detected with multiple flybys.}
    \label{fig:DeltaB}
\end{figure}

With Europa Clipper, multiple encounters with an altitude of 25 km at closest approach are planned \cite{campagnola2019tour}. For that purpose, we investigate the induction strength of a coupled system at 25 km in comparison to a radially symmetric system during a hypothetical flyby above the reservoir. Here, two effects that contribute to magnetic field perturbations in Europa's vicinity, additionally to induction by a subsurface ocean and reservoir, need to be considered, as they can potentially obscure the weak induction signal of a reservoir. The first is plasma interactions in Europa's ionosphere, which are largest when Europa is in the plasma sheet and can generate perturbations on the order of $100-200$ nT. Additional perturbations can be present in magnetometer measurements if the spacecraft crosses the Europan Alfvén wings \cite{schilling2007time, schilling2008influence}. The second phenomenon is apparently random small-scale fluctuations which occur on short time scales \cite{blocker2016europa}. \\ 
\change{To determine if induction signals of the reservoir can be distinguished from these fluctuations, we investigate reservoir signatures with various amplitudes at 25 km within the magnetometer measurements of the E14 flyby.}{To estimate the maximum amplitude that can be distinguished from these fluctuations, we superimpose several artificial amplitudes of reservoir induction signals onto the magnetometer data from the E14 flyby at an altitude of 25 km.} We specifically chose this encounter, as the fluctuations were small compared to other flybys. It is also one of the few Galileo orbits that are exempt from large-scale plasma effects. The E14 flyby also does not show any signs of plume activity, which can result in anomalies on the order of 100 nT \cite{jia2018evidence, arnold2019magnetic}. Thus, the E14 orbit offers clear identification of induction signals generated within Europa's subsurface. We added \add{artificial dipol signals as generated by a perfectly conducting} reservoir\remove{signals} with \add{various} amplitudes 5, 10, and 20 nT \add{at 25 km} to the measurements and found an induction response with a $5$ nT amplitude to be distinguishable from most fluctuations within the measurements, which lie at an average of $3$ nT (see Figure \ref{fig:E14}). However, individual fluctuations with amplitudes above 5 nT exist in the E14 data and occur on time scales similar to the induction response of a reservoir. Multiple flybys across the same region would be needed to verify such a weak perturbation due to induction from the reservoir, assuming the other small-scale fluctuations are random. An induction signal of 20 nT would be clearly distinguishable from fluctuations, which is also valid for an amplitude of 10 nT. The bottom panel of Figure \ref{fig:E14} displays a zoom-in of the time interval around the passage of the reservoir. Here, the dipole character of the reservoir's induction response is visible, which is different from the other fluctuations. This might additionally help to separate the apparently random fluctuations from perturbations caused by a reservoir. \add{In future measurements, where indications of an induced signal from a reservoir are present, a number of statistical tests based on the detailed structure of the RMS fluctuations will need to be applied to quantitatively assess such signals.} \\
In Figure \ref{fig:Flyby}, we present the radial component of the magnetic field $B_r$ that would be measured during a 25 km flyby above a reservoir with a radius of 20 km and a conductivity $\sigma = 30$ S/m. \add{The radial component $B_r$ is given as the sum of inducing field $B_{\mathrm{J},r}(t)$, given by equation} \ref{eq:Bfield}, \add{and induced field in ocean and reservoir $B_{\mathrm{ind},r}$, which is given by equation} \ref{eq:InductionResponse}. For comparison, we also include the radial component if no reservoir was present and if there was no coupling between ocean and reservoir. We see the induction signal of the reservoir as a small-scale perturbation overlaying the induction signal of the ocean. In this scenario, the coupling effects cause a small enhancement of the reservoir's induction response, resulting in a deviation of approximately $1$ nT compared to the induction response of a radially symmetric interior.  
We calculate the maximum deviation between these two models to obtain information about the potential detectability of local asymmetries within the parameter space discussed in section \ref{sec:pspace} (Figure \ref{fig:DeltaB}). For large reservoirs, its induction response ranges from $10^{-1}-10^{-2}$ nT. The induction signature decreases for smaller reservoirs down to orders of $10^{-4}$ nT. In the low conductivity limit, the induction response at 25 km lies below $10^{-2}$ nT for all radii. For all reservoir radii and conductivities considered in Figure \ref{fig:DeltaB}, the respective induction responses lie below 5 nT. \\
Based on Figure \ref{fig:DeltaB}, at 25 km closest approach, the magnetic field perturbation caused by a local reservoir of liquid water between the ocean and icy surface is too weak to be resolved, as the field strength lies below 5 nT across the entire parameter space considered in this study. Additionally, this induction response will likely not be distinguishable from overlaying plasma effects in the measurements described in \citeA{saur1998interaction}, 
\citeA{schilling2007time,schilling2008influence}, and \citeA{blocker2016europa}, which are generated by ionospheric interactions with the Jovian magnetosphere and perturb the magnetic field. As the values for all discussed induction responses from the reservoir lie below 5 nT, increasing the limit of detection to account for distinct identification of reservoir signals {has no impact on our conclusions}.

\subsection{Induced Fields at the Surface}
\begin{figure}
    \centering
    \includegraphics[width=0.9\textwidth]{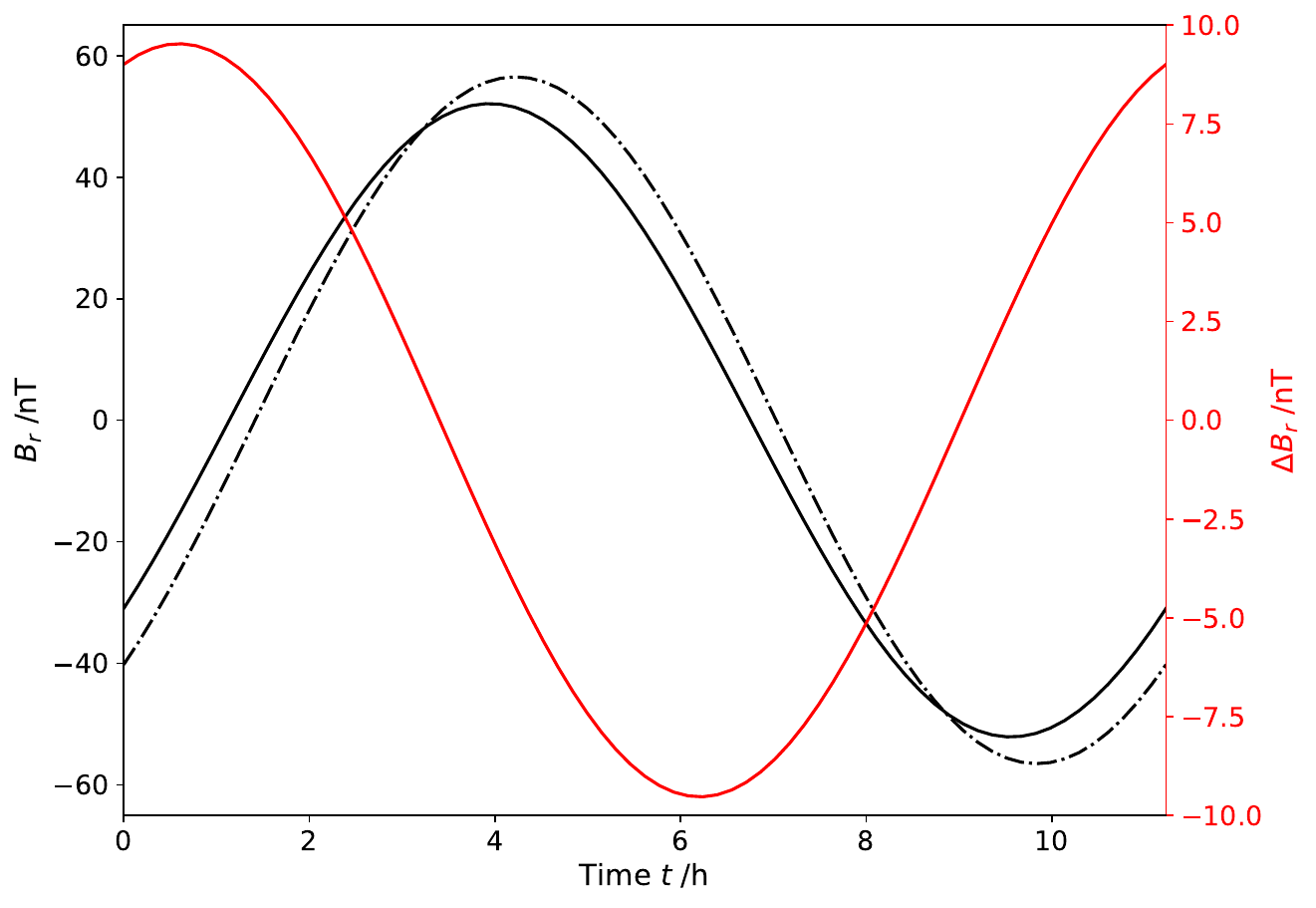}
    \caption{Time series of \add{the radial component of} the \add{sum of induced and inducing} magnetic field (solid\add{, black}) on top of a reservoir \remove{with} \add{and 1$^\circ$ in longitude outside a reservoir (dash dotted) for} $r_\res = 20$ km and $\sigma_\res $ = 30 S/m\remove{, (dash dotted) outside a reservoir.} The \add{solid} red \change{graph}{curve} and y-axis show the difference between the two calculations.}
    \label{fig:Timeseries}
\end{figure}

\begin{figure}
    \centering
    \includegraphics[width=0.9\textwidth]{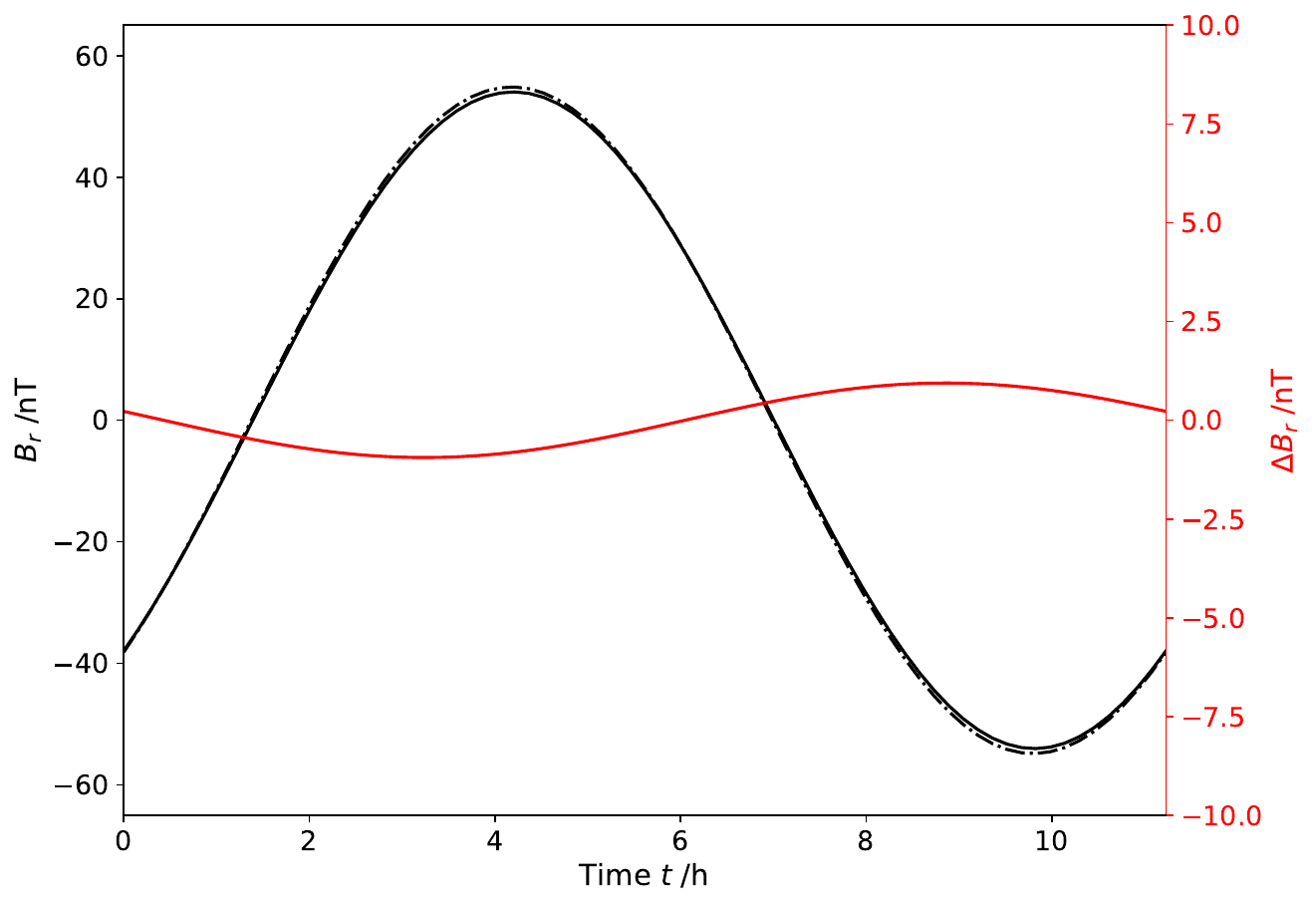}
    \caption{Same as Figure \ref{fig:Timeseries}, except that in this model only an ocean is present, i.e., there are no local asymmetries.}
    \label{fig:Timeseries_oc}
\end{figure}

\begin{figure}
    \centering
    \includegraphics[width=0.8\textwidth]{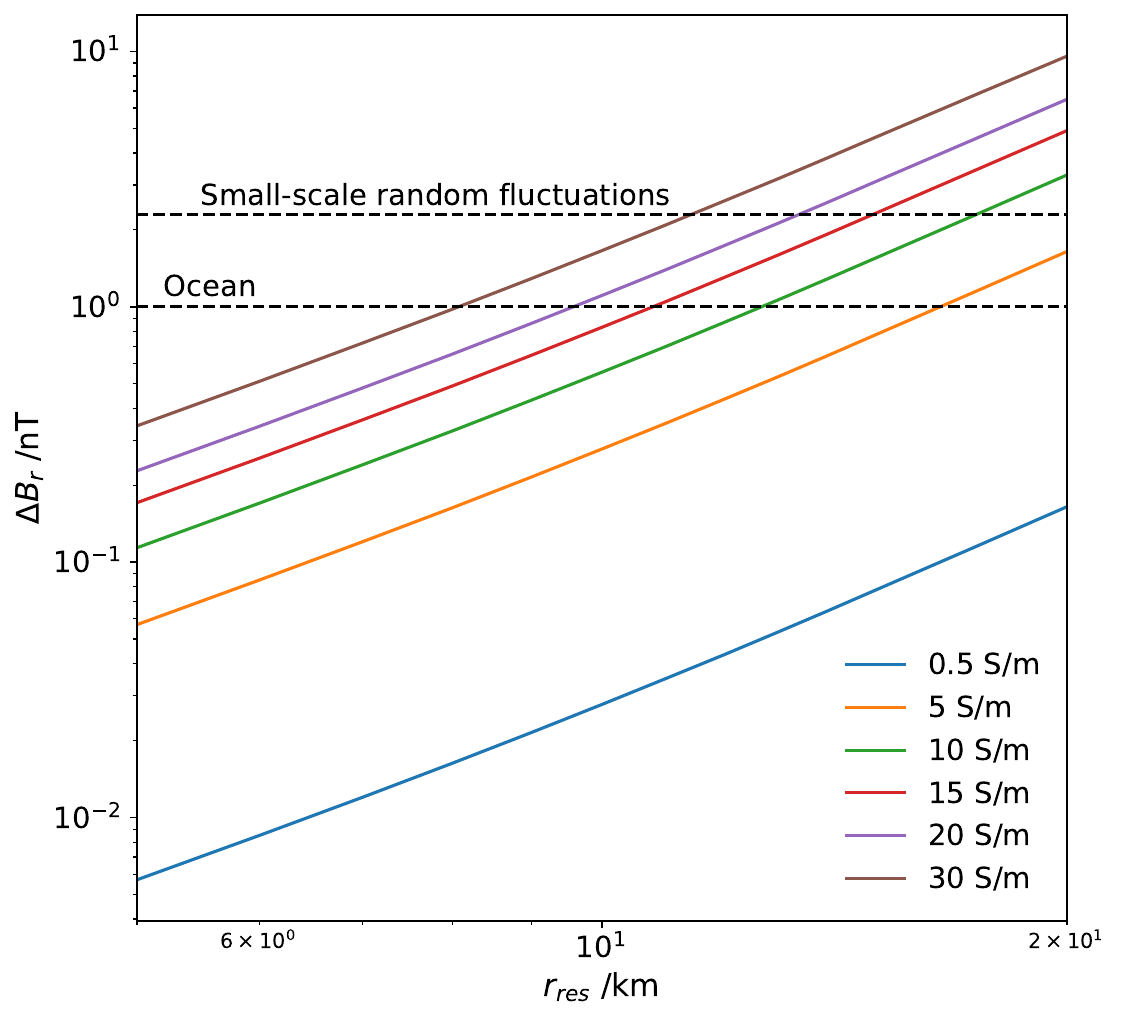}
    \caption{Maximum difference caused by the reservoir's induction response between surface measurements taken above a reservoir and \change{25 km}{1$^\circ$ in longitude (27.2 km)} away from the first magnetometer as a function of reservoir radius for conductivities ranging from 0.5 to 30 S/m. Here, the ocean's induction response has been subtracted from the difference. The dashed horizontal lines mark perturbations caused by the gradient over \change{25}{27.2} km distance of the ocean's induction response (1 nT) and the RMS at 2.3 nT from random noise in the data, respectively.}
    \label{fig:DeltaB_0km}
\end{figure}

Another potential option for the detection of local asymmetries could be the deployment of stationary magnetometers on Europa's surface, specifically on interesting targets such as chaos regions. This method requires the use of multiple magnetometers to obtain reference values outside the reservoir's reach, e.g., one magnetometer atop the reservoir and a second magnetometer which is positioned at a distance close to the reservoir, where the induction signal of the reservoir is approaching negligible values. For the results presented within this section, the first magnetometer is positioned at \change{$\theta=\phi=90^\circ$}{$\theta=90^\circ, \phi=0^\circ$} in Europa-centered coordinates, directly above the reservoir. The second magnetometer is stationed at \change{$\theta=90^\circ, \phi=91^\circ$}{$\theta=90^\circ, \phi=359^\circ$}, which corresponds to a distance of \change{approximately 25}{27.2} km between the two magnetometers. The use of magnetometers allows the recording of time series, which are shown in Figure \ref{fig:Timeseries}. As a comparison, we evaluate the different magnetic fields two magnetometers would measure if only an ocean is present (see Figure \ref{fig:Timeseries_oc}) and find that the deviation is smaller, with a maximum of 1 nT compared to approximately 9 nT if a reservoir with $\sigma = 30$ S/m and $r_\res = 20$ km is present.\\
While this difference in measurements of two nearby magnetometers primarily arises due to the rapid decrease of the reservoir's induction response, the spatial variation of the ocean's induction response contributes approximately up to 1 nT to the measured difference. In addition, the plasma effects can be large. We estimate the magnetic field gradient from Europa's plasma interaction based on simulations by \citeA{schilling2008influence} during the E4 flyby. While simulations made by other authors \cite <see e.g.,>[]{blocker2016europa,arnold2020plasma} would allow for such estimates as well, the model conditions assumed in \citeA{schilling2008influence} resemble our conditions the most due to the similarities between the E4 and E14 flyby, i.e., Galileo performed the flyby near the equatorial plane and Europa was well outside the plasma sheet. Near Europa's equatorial plane, the field varies by approximately 100 nT across one Europan radius, which for this flyby is mostly attributed to the induced field of the subsurface ocean. Thus, the plasma interactions appear to be smaller than induction effects outside the plasma sheet, with a gradient $< 1$ nT/km. Inside the plasma sheet, the gradient by plasma interactions is larger, up to approximately 200 nT/$R_\mathrm{E}$.\\ 
With the recording of temporal variation at a fixed position, random fluctuations will tend to average out over a long term, e.g., months of observations. To still account for such fluctuations in the measurements, we fit a polynomial of order 4 to 2 hours of E14 data outside the time region of closest approach and find a RMS error of 2.3 nT. However, during times in which Europa is within high density plasma regions, the magnetic field perturbations due to plasma interactions will make any induction signal unresolvable in the measurements. Outside the current sheet, we find that reservoirs with radius below 8 km cannot be detected for all conductivities considered and that reservoirs larger than 8 km require conductivities above $3-30$ S/m to lie above the gradient caused by the ocean's induction response (see Figure \ref{fig:DeltaB_0km}). For reservoirs below 12 km, the difference measured by two magnetometers is smaller than the RMS of the fluctuations in the data.\\
The three limits for the detectability of a reservoir due to gradients in the induction signal of the ocean, the plasma interaction and small-scale fluctuations discussed in the previous paragraph should be considered upper limits and smaller induction signatures of a reservoir are expected to be detectable. On long time series, periodic contributions can be filtered out and therefore suppressed within a statistically distributed noise, while the gradients due to the plasma interactions and the induced magnetic field from the ocean can be partially accounted for by numerical simulations. \\
\remove{To understand the dependencies of the reservoir's induction amplitude, it can be Taylor-approximated around small values for the argument of the Bessel function $r_\res k_\res \rightarrow 0$. This yields the following expression}
\remove{from which we can directly infer the importance of the reservoir's size when discussing its induction response and thus detectability. The induction amplitude increases with the square of the radius and linearly with both conductivity and rotation frequency of the inducing field.}

\section{Discussion and Conclusions} \label{SaC}
We briefly summarize our analytical model and afterward discuss the results presented in section \ref{Results}, especially with regard to the detectability of reservoirs with future missions.
\subsection{Analytical Model for Coupled Induction}
We constructed an analytical model that accurately describes the electromagnetic coupling and the resulting induced fields between two neighboring spherically symmetric bodies, in this case a global subsurface ocean and a small-scale reservoir within the icy crust. \add{This model is solved numerically using an iterative method to account for the mutual induction between ocean and reservoir.} The implementation of coupling effects is crucial for the physical correctness of the calculated fields, and our work provides a new approach to describing these coupled fields. The equations used in this approach assume both conductors to be radially symmetric with a constant conductivity. It is important to consider each multipole moment of the induced field separately as the phase shift is a function of degree $l$, thus each contribution to the multipole is induced with a varying temporal delay $\phi^\mathrm{ph}_l/\omega$. \add{It should be noted that real reservoirs or local conductivity anomalies will not be perfectly spherically symmetric. For complex and arbitrarily shaped conductive structures, no analytical solution of the induction response exists, and the induction response needs to be calculated with full numerical solvers. Such calculations are outside the scope of this work.}
   
\subsection{Detectability during a 25 km Flyby} \label{sec:Det25km}
The detectability of a small-scale reservoir \change{is facing}{faces} multiple challenges. First, the low amplitude of a reservoir due to its small size, as the induction amplitude varies with the square of the radius via equation \ref{eq:Ares}. For a reservoir with radius $r_\res = 20$ km and conductivity $\sigma=30$ S/m, its induction amplitude is still just $A_\res = 0.15$, while for an ocean such large conductivities would result in a near-perfectly conducting response. If we now assume a perfectly conducting reservoir with $A=1$ to test its detectability at 25 km, this illustrative scenario highlights the challenge imposed by the coupling processes between ocean and reservoir. Here, the reservoir's dipole response to the Jovian background field gets obscured by the coupled induction response to the ocean's dipole, resulting in a significantly lower induction signature by the reservoir. At 25 km altitude, one would measure a deviation to the symmetric model of approximately 1 nT after coupling processes instead of 8 nT if only the superposition is assumed \add{(superposition refers to considering just the $n=1$ iteration for both ocean and reservoir)}. The third issue is the distance to the reservoir, as its induced fields experience a rapid decrease due to its small extension. These fields are dipole dominant and decrease with $(r_\res/r)^3$, where $r>2 r_\res$ for our cases.  \\
A reservoir might be detectable if its properties lie outside our considered range, i.e., if the reservoir has a higher conductivity or a radius larger than 20 km. A conductivity above the upper limit of 30 S/m could hint at a higher temperature or pressure within the water melt, resulting in an increased conductivity of saline water containing sodium chloride \cite{guo2019electrical}. Although past studies argued for a magnesium sulfate rich ocean \cite{kargel2000europa, mckinnon2003sulfate}, its conductivities are \remove{, in general,} lower than sodium chloride \add{assuming near-saturation conditions} \cite{hand2007empirical}, and would thus generate weaker magnetic field perturbations. A larger radius will be beneficial for its detectability due to a larger induction amplitude and the slower decrease of the resulting induced field. However, this also implies a deeper ocean, as the icy crust thickness has to be adjusted accordingly. As a deep subsurface ocean lies outside of many of the estimates presented in section \ref{sec:iceshell}, we chose not to consider such cases in our study.

\subsection{Detectability at the Surface}
A reservoir can be detected if magnetometers are deployed on Europa's surface, which would be \change{an exciting experiment}{a valuable experiment to include} in future missions. For that, one magnetometer would be positioned directly on the region of interest, for example a chaos region. A second magnetometer is placed right outside the chaos region, where the induction signal of a local water melt below it would approach negligible values. This method allows distinguishing local features from the global depth variabilities of the ocean considered by \citeA{styczinski2022perturbation}. While surface measurements also eliminate one of the challenges mentioned in section \ref{sec:Det25km} in detecting a reservoir, i.e., the rapid decrease of the induction response with distance, the induction amplitudes remain limited to low values. Thus, in our most conservative case, only reservoirs with a radius larger than $r_\res \approx 12$ km and conductivities $\sigma > 7$ S/m are able to generate an induction response where the difference measured by two magnetometers at \change{25}{27.2} km distance lies above the RMS inferred from the data at times around Galileo's E14 flyby. The recording of temporal variability at a fixed position could, however, allow us to obtain a more profound understanding of the periodicity of the various electromagnetic effects and distinguish them from each other in the time series. With this information, large-scale plasma effects can potentially be subtracted from measurements using numerical models, which would then allow for detection of reservoir signals even in larger\add{-}plasma\add{-}density regions. Model calculations, where the reservoir placements resemble those of chaos regions on Europa, can be made to compare future measurements with the theoretical results. 
\subsection{Outlook to Future Spacecraft and Lander Missions}
Finally, we will discuss our results in the context of the planned Europa Clipper and JUICE spacecraft, as well as highlight the advantages of a surface lander.
\subsubsection{Detectability with JUICE}
For the JUICE mission, only two flybys at Europa are planned, with altitudes around 400 km \cite{cappuccio2022callisto}. Assuming a reservoir with a radius of 20 km, its induced field decreases to about \change{1/64000}{1/8000} of its amplitude at the planned altitude, assuming a simple dipole\add{, equaling to $4\cdot10^{-3}$ nT for $\sigma_\res = 30$ S/m}. Thus, JUICE will not be able to detect reservoirs of liquid water with its magnetometer measurements.
\subsubsection{Detectability with Europa Clipper}
In this work, Europa Clipper has served as a motivation in choosing an altitude of 25 km during a hypothetical flyby, as multiple encounters at that distance are planned. However, assuming a sensitivity limit of 5 nT, the induction signals of a reservoir are too weak to be resolved with magnetometer measurements. If the measurements would have significantly smaller fluctuations than those measured during the E14 flyby, then the sensitivity could be lower than 5 nT as well. This potentially allows the detection of large and highly conductive reservoirs.
\subsubsection{Detectability with Landers}
\change{Reservoirs are shown to be}{We have shown that reservoirs are expected to be} detectable at the surface if Europa is outside the plasma sheet, providing \change{an exciting outlook to}{additional scientific capabilities to consider for} future plans for lander missions, which have been the subject of discussion over the last years \cite <see e.g.,>[]{pappalardo2013science,blanc2020joint,hand2022science}. Additionally, the idea of a magnetometer network across Europa's surface offers an interesting approach to map spatial variabilities, e.g., depth variabilities between the polar and equatorial regions. Such a network of surface magnetometers would be highly valuable to investigate multiple chaos regions suspect to the existence of water pockets and help in the understanding of their formation.

\subsection{Explanation of Spherical Assumption}
In this work, we investigated the joint induction response of two spherically symmetric, electrically conducting bodies, i.e., an ocean and a reservoir. \add{We provided an analytical model that is solved using a numerical, iterative method and calculated the total induction response of the coupled system across the chosen parameter space for the reservoir's radius and conductivity. These induced magnetic fields have been investigated for their detectability at altitudes equal to the planned JUICE and Europa Clipper spacecraft flybys, where for Europa Clipper the minimum flyby altitude of 25 km has been chosen. With the JUICE spacecraft, reservoirs cannot be detected with magnetometer measurements, whereas Europa Clipper might be able to resolve large and highly conductive reservoirs assuming small random fluctuations at the time of the flybys. Liquid water reservoirs are likely detectable at Europa's surface using a network of magnetometers, offering a valuable outlook to future lander missions.} \remove{Real reservoirs or local conductivity anomalies, however, will not be perfectly spherically symmetric. For complex and arbitrarily shaped conductive structures, no analytical solution of the induction response exists, and the induction response needs to be calculated with full numerical solvers. Such calculations are outside the scope of this work.} The \add{presented} study \remove{here} is therefore a first step towards understanding the induction response and the detectability of local near-surface anomalies within Europa's icy crust by induction sounding.

\appendix
\section{Coordinate Transformation}\label{app:Transform}
\begin{figure}
    \centering
    \includegraphics[width=0.9\textwidth]{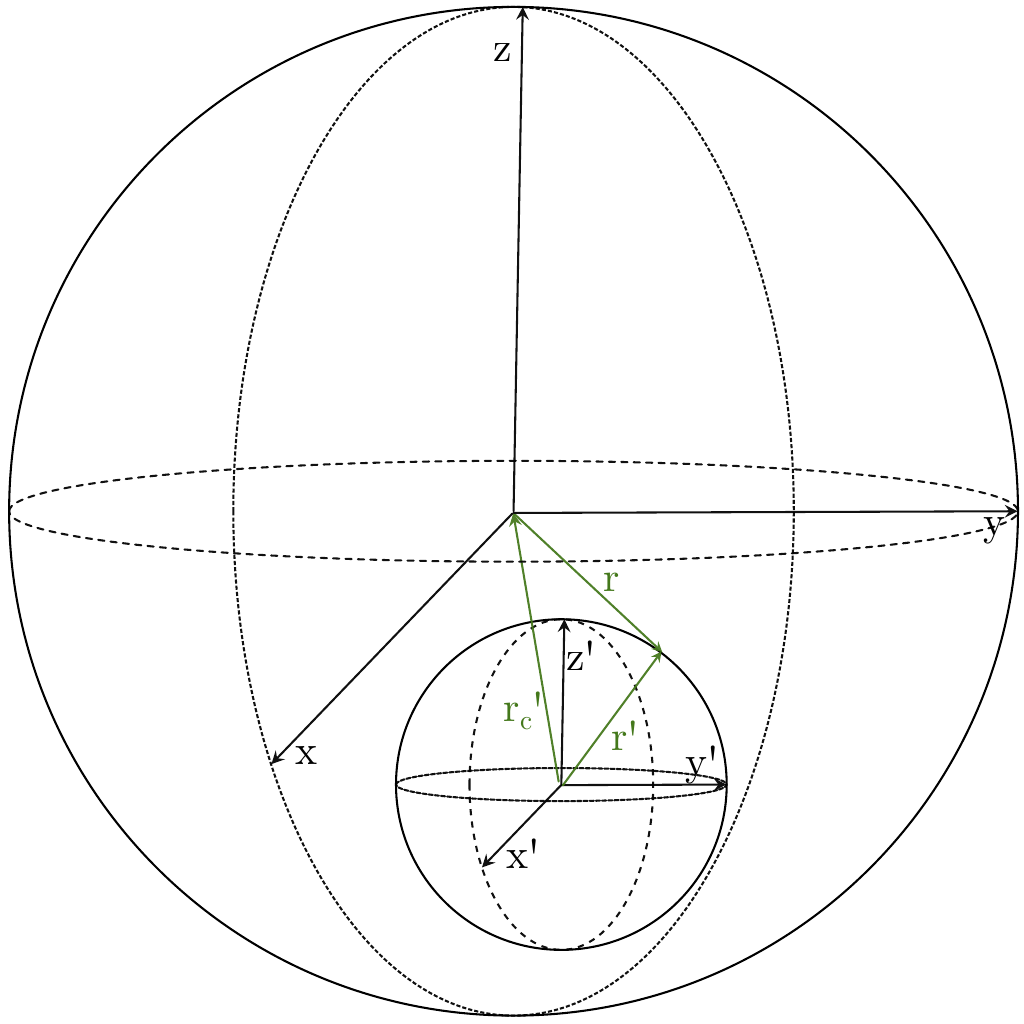}
    \caption{Sketch of the two coordinate systems with origin in Europa's center $(x,y,z)$ and with origin in the reservoir $(x',y',z')$, respectively. The scale and position of the reservoir are chosen arbitrarily. The green vectors visualize the transformation between the two systems.}
    \label{fig:Trafo}
\end{figure}
The calculation of Gauss coefficients requires knowledge of the external field on the surface of the conducting body. For that, we need to transform the ocean's induction response into reservoir-centered coordinates and vice versa (see Figure \ref{fig:Trafo}).
We start with two spherical coordinate systems $(r,\theta, \phi)$ for the ocean and $(r', \theta', \phi')$ for the reservoir respectively and introduce the transformation vector $\vecr_\mathrm{c} = (r_{\mathrm{c},x},r_{\mathrm{c},y},r_{\mathrm{c},z})$, in the cases presented here $\vecr_\mathrm{c} = (r_\mathrm{c}, 0, 0)$.
In ocean-centered Cartesian \change{components}{coordinates}, we express the surface of the reservoir as
\begin{linenomath*}
\begin{align*}
    x &= r_\mathrm{c} + r_\res \sin \theta \cos \phi \\
    y &= r_\res \sin \theta \sin \phi \\
    z &= r_\res \cos \theta
\end{align*}
\end{linenomath*}
and 
\begin{linenomath*}
\begin{align*}
    x' &= -r_\mathrm{c} + r_0 \sin \theta' \cos \phi' \\
    y' &= r_0 \sin \theta' \sin \phi' \\
    z' &= r_0 \cos \theta'
\end{align*}
\end{linenomath*}
for the ocean's surface in reservoir-Cartesian coordinates. After transforming the Cartesian coordinates into spherical coordinates via
\begin{linenomath*}
\begin{align*}
    r &= \sqrt{x^2 + y^2 + z^2} \\
    \theta &= \atantwo(\sqrt{x^2+y^2}, z) \\
    \phi &= \atantwo(y,x),
\end{align*}
\end{linenomath*}
we compute the radial component $B_r(r,\theta,\phi)$. The Gaussian method, however, requires us to transform the radial component $B_r(r,\theta,\phi) \rightarrow B_{r'}(r,\theta,\phi)$ for correct application, which is done by transforming into reservoir-centered Cartesian \change{components}{coordinates}
\begin{linenomath*}
\begin{align*}
    B_{x'} &= B_r \sin \theta \cos \phi + B_\theta \cos \theta \cos \phi - B_\phi \sin \phi \\
    B_{y'} &= B_r \sin \theta \sin \phi + B_\theta \cos \theta \sin \phi + B_\phi \cos \phi \\
    B_{z'} &= B_r \cos \theta - B_\theta \sin \theta
\end{align*}
\end{linenomath*}
and back into spherical
\begin{linenomath*}
\begin{equation*}
    B_{r'} = B_{x'} \sin \theta' \cos \phi' + B_{y'} \sin \theta' \sin \phi' + B_{z'} \cos \theta'.
\end{equation*}
\end{linenomath*}
The transformed radial component is used to calculate the external Gauss coefficients in equation \ref{eq:gext}.
%%%%%%%%%%%%%%%%%%%%%%%%%%%%%%%%%%%%%%%%%%%%%%%
% Optional Glossary, Notation or Acronym section goes here:
%
% Glossary is only allowed in Reviews of Geophysics
%  \begin{glossary}
%  \term{Term}
%   Term Definition here
%  \term{Term}
%   Term Definition here
%  \term{Term}
%   Term Definition here
%  \end{glossary}

%%%%%%%%%%%%%%%%%%%%%%%%%%%%%%%%%%%%%%%%%%%%%%%
% Acronyms
%% NOTE that acronyms in the final published version will be spelled out when used in figure captions.
%   \begin{acronyms}
%   \acro{Acronym}
%   Definition here
%   \acro{EMOS}
%   Ensemble model output statistics
%   \acro{ECMWF}
%   Centre for Medium-Range Weather Forecasts
%   \end{acronyms}

%%%%%%%%%%%%%%%%%%%%%%%%%%%%%%%%%%%%%%%%%%%%%%%
% Notation
%   \begin{notation}
%   \notation{$a+b$} Notation Definition here
%   \notation{$e=mc^2$}
%   Equation in German-born physicist Albert Einstein's theory of special
%  relativity that showed that the increased relativistic mass ($m$) of a
%  body comes from the energy of motion of the body—that is, its kinetic
%  energy ($E$)—divided by the speed of light squared ($c^2$).
%   \end{notation}

%%%%%%%%%%%%%%%%%%%%%%%%%%%%%%%%%%%%%%%%%%%%%%%
%
% DATA SECTION and ACKNOWLEDGMENTS
%
%%%%%%%%%%%%%%%%%%%%%%%%%%%%%%%%%%%%%%%%%%%%%%%
\newpage
\section*{Data Availability Statement}
This work used magnetometer measurements of the Galileo spacecraft, which were archived by the NASA Planetary Data System: Planetary Plasma Interaction \add{and can be found here https://pds-ppi.igpp.ucla.edu/search/view/?f=yes\&id=pds://PPI/galileo-mag-jup-calibrated/data-highres-europa/ORB14\_EUR\_EPHIO\&o=1} \cite{pds3mag}. The model has been implemented with Python, with which the results were obtained as well. The source code is publicly available in a repository \cite{software}.

%This section MUST contain a statement that describes where the data supporting the conclusions can be obtained. Data cannot be listed as ''Available from authors'' or stored solely in supporting information. Citations to archived data should be included in your reference list. Wiley will publish it as a separate section on the paper’s page. Examples and complete information are here:
%https://www.agu.org/Publish with AGU/Publish/Author Resources/Data for Authors

\acknowledgments
This project has received funding from the European Research Council (ERC) under the European Union's Horizon 2020 research and innovation programme (grant agreement No. 884711).

%%%%%%%%%%%%%%%%%%%%%%%%%%%%%%%%%%%%%%%%%%%%%%%
% REFERENCES and BIBLIOGRAPHY
%
% \bibliography{<name of your .bib file>} don't specify the file extension
% don't specify bibliographystyle
%
%%%%%%%%%%%%%%%%%%%%%%%%%%%%%%%%%%%%%%%%%%%%%%%

\bibliography{bibliography}

%Reference citation instructions and examples:
%
% Please use ONLY \cite and \citeA for reference citations.
% \cite for parenthetical references
% ...as shown in recent studies (Simpson et al., 2019)
% \citeA for in-text citations
% ...Simpson et al. (2019) have shown...
%
%
%...as shown by \citeA{jskilby}.
%...as shown by \citeA{lewin76}, \citeA{carson86}, \citeA{bartoldy02}, and \citeA{rinaldi03}.
%...has been shown \cite{jskilbye}.
%...has been shown \cite{lewin76,carson86,bartoldy02,rinaldi03}.
%... \cite <i.e.>[]{lewin76,carson86,bartoldy02,rinaldi03}.
%...has been shown by \cite <e.g.,>[and others]{lewin76}.
%
% apacite uses < > for prenotes and [ ] for postnotes
% DO NOT use other cite commands (e.g., \citet, \citep, \citeyear, \nocite, \citealp, etc.).
%

\end{document}